\pdfoutput=1

\documentclass[11pt]{article}

\usepackage{authblk}

\usepackage[]{ACL2023}

\usepackage{times}
\usepackage{latexsym}
\usepackage{graphicx}
\usepackage{tabularx}
\usepackage{subfig}
\usepackage{comment}
\usepackage[T1]{fontenc}

\usepackage[utf8]{inputenc}

\usepackage{microtype}

\usepackage{inconsolata}
\usepackage{array}
\usepackage{bm}
\usepackage{float}

\usepackage{stfloats}

\usepackage{times}
\usepackage{latexsym}
\usepackage{hyperref}

\newcommand{\RNum}[1]{\uppercase\expandafter{\romannumeral #1\relax}}

\usepackage{multirow}
\usepackage{tcolorbox}
\usepackage{tabularx}
\usepackage{makecell}
\usepackage{amsmath,amssymb}
\usepackage{physics}
\usepackage{booktabs}
\usepackage{enumitem}
\usepackage{subcaption}
\usepackage[switch]{lineno}
\usepackage{listings}
\usepackage{xcolor}

\definecolor{codegreen}{rgb}{0,0.6,0}
\definecolor{codegray}{rgb}{0.5,0.5,0.5}
\definecolor{codepurple}{rgb}{0.58,0,0.82}
\definecolor{backcolour}{rgb}{0.95,0.95,0.92}

\lstdefinestyle{mystyle}{  
    commentstyle=\color{codegreen},
    keywordstyle=\color{magenta},
    numberstyle=\tiny\color{codegray},
    basicstyle=\ttfamily\small,
    breakatwhitespace=false,         
    breaklines=true,                 
    captionpos=b,                    
    keepspaces=true,                        
    numbersep=1pt,                  
    showspaces=false,                
    showstringspaces=false,
    showtabs=true,                  
    tabsize=2
}

\newcommand{\tool} {{\tt DocCGen}}
\newcommand{\nlyaml} {{\tt NL to Ansible-YAML}}
\newcommand{\tldr} {{\tt TLDR}}
\newcommand{\nlbash} {{\tt NL2Bash}}

\newcommand{\docp} {{\tt DocPrompting}}
\newcommand{\ansbileaware} {Ansible Aware}
\newcommand{\schemacorrect} {Schema Correct}

\newcommand{\modulematch} {Module Acc}
\newcommand{\nltocode}{NL-to-Code} 

\title{DocCGen: Document-based Controlled Code Generation}

\author[a]{\bf Sameer Pimparkhede}
\author[b]{\bf Mehant Kammakomati}
\author[b]{\bf Srikanth G. Tamilselvam} 
\author[b]{\\ \bf Prince Kumar}
\author[b]{\bf Ashok Pon Kumar}
\author[a]{\bf Pushpak Bhattacharyya}

\affil[a]{IIT Bombay}
\affil[b]{IBM Research}
\affil[a]{\{sameerp,pb\}@cse.iitb.ac.in}
\affil[b]{{\{mehant.kammakomati2,prince.kumar12\}@ibm.com}}
\affil[b]{{\{srikanth.tamilselvam,ashokponkumar\}@in.ibm.com}}

\begin{document}
\maketitle
\begin{abstract}


Recent developments show that Large Language Models (LLMs) produce state-of-the-art performance on natural language (NL) to code generation for resource-rich general-purpose languages like C++, Java, and Python. However, their practical usage for structured domain-specific languages (DSLs) such as YAML, JSON is limited due to domain-specific schema, grammar, and customizations generally unseen by LLMs during pre-training. Efforts have been made to mitigate this challenge via in-context learning through relevant examples or by fine-tuning. However, it suffers from problems, such as limited DSL samples and prompt sensitivity 
but enterprises maintain good documentation of the DSLs. 
Therefore, we propose \tool{}, a framework that can leverage such rich knowledge by breaking the \nltocode{} generation task for structured code languages into a two-step process. First, it detects the correct libraries using the library documentation that best matches the NL query. Then, it utilizes schema rules extracted from the documentation of these libraries to constrain the decoding.
We evaluate our framework for two complex structured languages, Ansible YAML and Bash command, consisting of two settings: Out-of-domain (OOD) and In-domain (ID). Our extensive experiments show that \tool{} consistently improves different-sized language models across all six evaluation metrics, reducing syntactic and semantic errors in structured code\footnote{We plan to open-source the datasets and code to motivate research in constrained code generation.}.
\end{abstract}

\begin{figure}[ht]
 \includegraphics[width=0.5\textwidth]{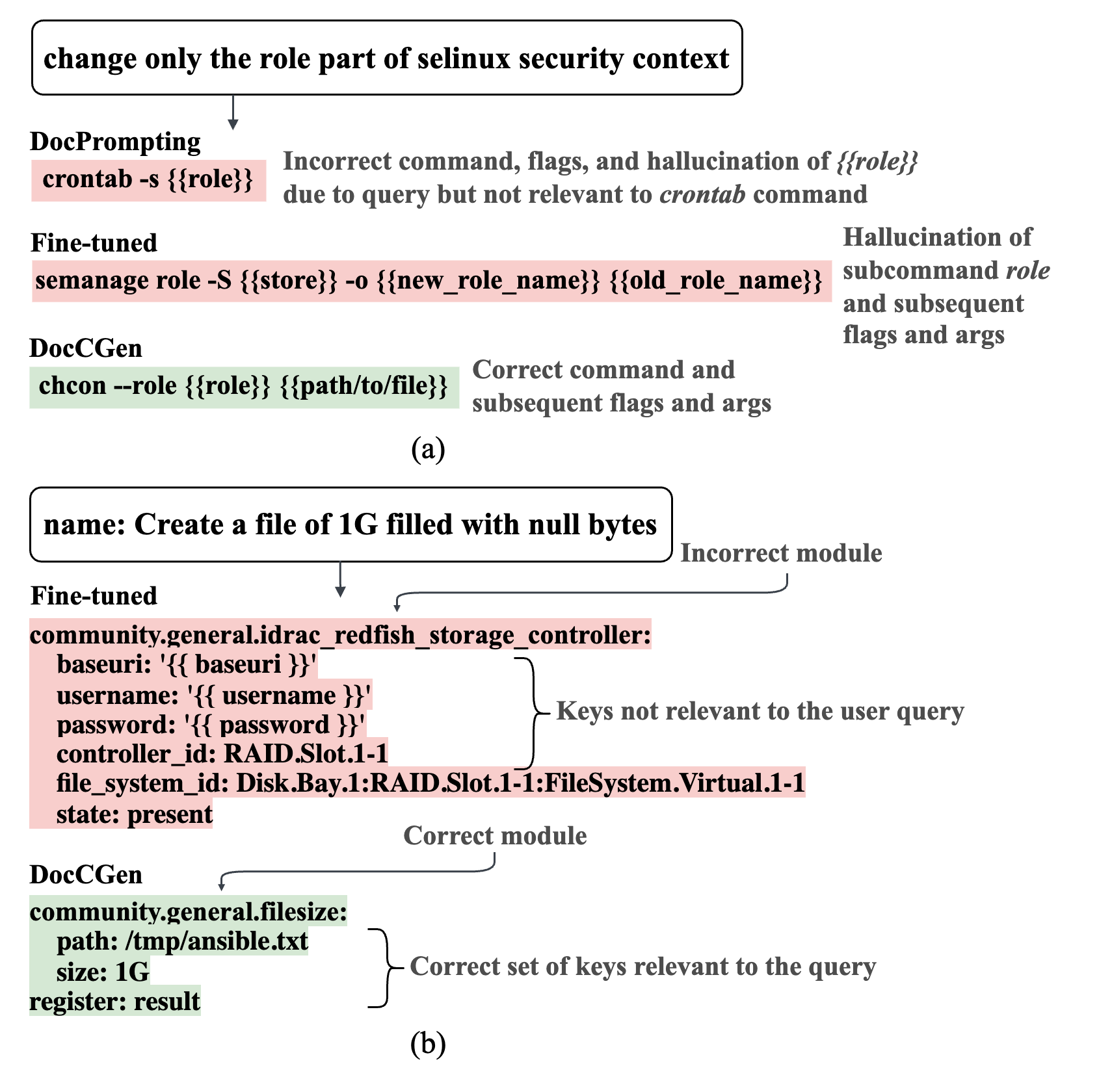}
\caption{Illustration of shortcomings with fine-tuning and DocPrompting \citep{zhou2022doccoder} approaches with an example for (a) NL to Bash task (uses GPT Neo 1.3B) and (b) \nlyaml{} task (uses StarCoder2 3B) and the proposed DocCGen method to overcome the limitations.}
\label{Example figure}
\end{figure}

\section{Introduction}\label{Introduction}
\begin{figure*}
\centering
 \includegraphics[width=1\textwidth]{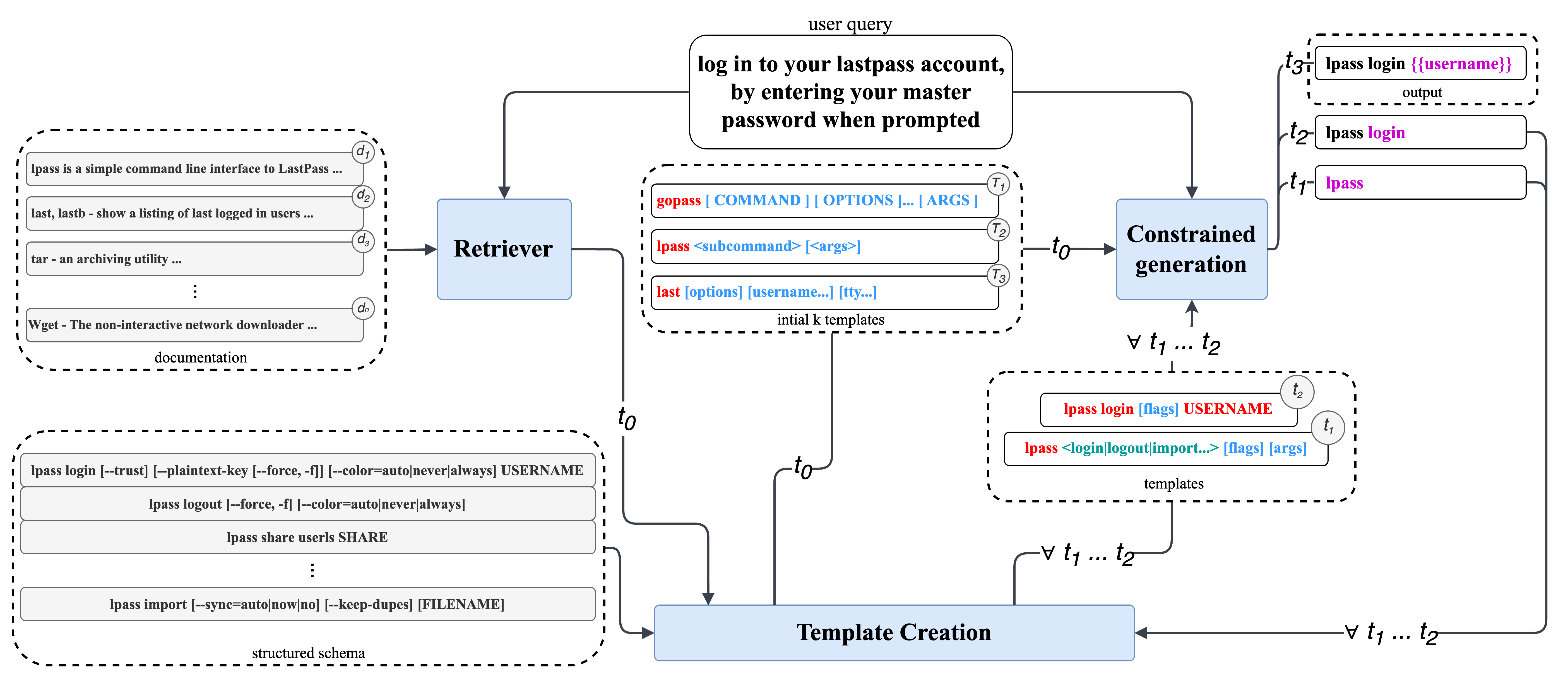}
  \caption{Overview of \tool{}. For a given user query, top $k$ relevant library documentations are retrieved and for which initial $k$ templates are created. \emph{Static} part of the template is shown in red, while the \emph{variable} part is in blue. The variable field with a fixed position in the code is enclosed in angle brackets, for instance <subcommand>, as shown in the initial k templates block in the figure. The model is guided to follow one of the templates during decoding. Each time step $t_i$ shows the step-by-step dynamic template evolution and constrained decoding output, adhering to the time-step template leading to the final generated code at $t3$.}
\label{Figure 1}
\end{figure*}

The Natural Language to Code (\nltocode{}) task has become pivotal in the intersection of natural language processing and programming.
\nltocode{} systems can help engineers write a program efficiently by conveying their intentions at a higher level, as shown in Figure \ref{Example figure}. 
Systems like Amazon code Whisperer\footnote{\url{https://aws.amazon.com/codewhisperer/}}, GitHub Co-pilot\footnote{\url{https://github.com/features/copilot/}} perform well in \nltocode{} task due to large language models (LLM) trained on extensive data. While they perform well in general resource-rich languages like C++, Python, or Java, their practical usage in structured DSL is limited. DSLs are enterprise-specific languages with specialized schemas and syntax suitable for a specific domain or application\footnote{\url{https://w.wiki/6jCH}}. Numerous enterprises use structured languages like Bash, YAML, JSON and HCL (HashiCorp Configuration Language) with specific customizations for automation and to configure and manage infrastructure in IT environments. These languages or their customizations are potentially unseen by LMs during pre-training, limiting their practical usage \citep{zan-etal-2022-language}. Some existing methods attempt to address this challenge via in-context learning through examples \citep{poesia2022synchromesh}, by fine-tuning \citep{10247987} or by using relevant documentation as additional context \citep{zan-etal-2022-language, zhou2022doccoder, parvez2021retrieval, lu2022reacc}. However, relevant context or samples available for DSL are often insufficient to incorporate diverse library schema rules or specialized structure knowledge in the LM \citep{zan-etal-2022-language, wang2024grammar}. This results in hallucination and different syntactic and semantic errors, as shown in Figure \ref{Example figure}.
However, enterprises usually maintain detailed documentation of their custom libraries (e.g. ansible modules, bash utilities), including the descriptions, schema, and syntax, to assist developers in enforcing structure and maintaining data integrity. We believe such schema and documentation can be better leveraged during code generation. Therefore, we propose a framework \tool{} that treats the \nltocode{} task as a two-step process, each heavily relying on the documentation. The first step identifies relevant code libraries for the task by retrieving the library documentation relevant to the NL query. 
The second step employs constrained decoding (CD) to guide code generation by using the grammar and schema rules extracted from the documentation of libraries identified in the first step, as shown in Figure \ref{Figure 1}.
We evaluate this approach for two diverse and complex structured languages, Ansible YAML and Bash command. Generation for these languages is tricky due to complexities like the diverse library schemas, optional and required fields, the order-agnostic nature of fields, and inter-field dependencies. We believe studying these complex structures encompasses most of the challenges in other structured DSLs and allows easily extending \tool{} to other domains. 
Since the major challenge in DSLs is the limited availability of samples, we focus on enhancing performance for unseen code libraries or libraries with very few samples in the training corpus. Hence, we evaluate our approach in two settings: In-domain and Out-of-domain. Similar to \citet{zhou2022doccoder}, none of the libraries in the test set are seen during training in the OOD setting. In the ID setting, every library in the test set has very few \nltocode{} pairs in the train set.
\tool{} consistently improve over state-of-the-art models and techniques by a significant margin (Table \ref{table:OOD}, \ref{table:In-domain}) across multiple settings.

Finally, we introduce first \emph{publicly} available benchmark dataset for NL to structured code generation task consisting of Ansible-YAML language. Intricate challenges in Ansible-YAML generation, like the complex structure and diverse module schemas, lead to subpar performance even for fine-tuned code LMs (Table \ref{table:OOD}). We curate \nlyaml{} dataset with $18k$ samples with code snippets from more than $2500$ modules under OOD and ID settings (Table \ref{table:data-stats}). More information and examples for Ansible-YAML are presented in section \ref{subsec:appendix-ansibleyaml}.
Besides this, we augment new \nlyaml{} and existing NL to Bash dataset \tldr{} \citep{zhou2022doccoder} with descriptions, detailed schema and grammar information from each library. We believe these datasets will advance research in constrained generation and handling low-resource or unseen data scenarios in structured DSLs.

\textbf{Our contributions are:}
\begin{enumerate}
    \item A novel framework that treats the NL to structured code generation task as a two-step process. While the first step detects the correct code libraries for the task, the second step employs constrained decoding to enforce schema adherence based on the schema rules extracted from the documentation. 
    \item An extensive study on two diverse structured languages, Bash command and Ansible YAML, for Out-of-domain and In-domain settings. The results show our framework outperforms state-of-the-art techniques across all six metrics (Table \ref{table:OOD}, \ref{table:In-domain}) for different-sized models.
   \item New datasets a) \nlyaml{} dataset with $18k$ pairs (refer to Table \ref{table:data-stats}). b) Descriptions and schema of Ansible YAML modules and bash utilities (Section \ref{Dataset}) to further motivate research in DSL code generation.
    
\end{enumerate}

\section{Related Work}

\noindent\textbf{Constrained decoding:} Controlled code generation using constraints has been previously studied majorly for the text-to-SQL task, using plan-based static templates \citep{bhaskar2023benchmarking} or SQL parser-based semantic checks \citep{scholak-etal-2021-picard}. The database schema is fixed and given as input with a text query for text-to-SQL. However, we target a more complex problem involving multiple libraries and diverse schemas and use library documentation to solve this.
\citet{poesia2022synchromesh} and \citet{wang2024grammar} use in-context learning via relevant samples or grammar strings and constrain the decoding further. However, in-context learning does not solve the issue of the correctness of the library. Hence, we instead follow a two-step process using library documentation.
\citet{agrawal2023guiding} uses constrained decoding for general-purpose languages like Java and C\# using suggestions from intelligent parsers. However, such advanced parsers are uncommon for DSLs and might provide incomplete constraints. Hence, we use rules extracted from documentation more commonly available.

\noindent\textbf{Context Based Controlled Generation like RAGs:}
     Many existing methods retrieve the relevant context and augment it with the input prompt to improve the code generation \citep{lu2022reacc, zan-etal-2022-language, zhou2022doccoder, parvez2021retrieval, ding2022cocomic}. Although effective, these methods do not ensure schema and grammar adherence, especially for unseen libraries and languages. \citet{zhang2023repocoder} and \citet{zan-etal-2022-language} improve over vanilla retrieval-augmented code generation but require either architectural changes or extra pre-training. Hence, unlike these methods, we guide the generation by adjusting the output logits.
    

\section{DocCGen Framework}


\tool{} is a two-stage framework: The first stage uses information retrieval (IR) to detect relevant libraries. The second stage uses the neuro-symbolic constrained decoding to control generation and ensure adherence to the schema of relevant libraries.

\subsection{Background and Definitions}
For a given NL query $q$, we generate a code snippet $c$. The first stage of the framework uses a set of documentation $D$, collected using library descriptions as described in section \ref{Dataset}. Hence, each document in $D$ describes the respective library. In this section, we define some frequently used terms.

\paragraph{Structured schema:} Structured schema stores the list of valid keywords for every field and the inter-field dependency information. For example, the structured schema of any bash utility (e.g., \emph{cat} or \emph{tar}) includes information like a list of optional and required sub-commands, flags, and inter-field dependency information (e.g., a list of valid flags and arguments for a sub-command).
\paragraph{Template:} The template encodes the structure of the code snippet for the library as a string and is used to guide the model during decoding. While the structured schema maintains a list of valid keywords for every field, the template encodes the positional information of fields in the code snippet. Every template has a \emph{static} and \emph{variable} part. The static part is directly copied in the output code, and the model generates the variable part adhering to the library schema. For Ansible YAML and bash, the template starts with the static part, typically the library name or its variation used in actual code.
For example, for the bash utility \emph{git-mv}, template is \emph{git mv [options] \{\{source\}\} \{\{destination\}\}}. In this template, \emph{[options]} is a variable part and represents the sequence of flags in the command to be generated by the model. The other part is static and is directly included in the output code. Structured schema and template together represent the grammar of the library in the format, which can be easily used to guide the decoding. More example templates are presented in the listing \ref{lst:bashsyntax}.

\paragraph{Trigger signals:} Trigger signals $G$ comprises rules to control the generation of optional fields (fields with context-dependent presence and positions) or conditions to dynamically change the template. When triggered, the guiding template changes and makes the model follow new specified rules. For example, generating the " --" token in bash triggers valid doublehand flag generation or generation of pipe operator (token "|") triggers the start of a new process enabling to control generation of command with multiple bash utilities. In YAML, indentation beyond the first level triggers the generation of nested schema with completely different rules from the parent schema, forming a new guiding template.
Details of all triggers can be found at \ref{Bash triggers} and \ref{YAML triggers}.

\noindent\subsection{Framework}
For the given NL query $q$, the first stage of the framework retrieves $k$ most relevant documents $D^*$ from a pool of documents $D$. This gives us a set of $k$ most relevant libraries that can be used to generate code $c$. Then, we fetch the initial templates of every retrieved library stored offline. The next step instantiates the generator model to generate the code snippet $c$. During auto-regressive inference decoding, the model is constrained to follow one of the $k$ code templates. As the decoding proceeds, the template might be changed dynamically based on the tokens generated by the model, the structured schema of the library, and trigger signals, as shown in Figure \ref{Figure 1}.


\subsection{Information retrieval}
We experiment with sparse and dense retrieval systems in the first stage of \tool{}.

\subsubsection{Sparse retrieval}
We use the BM25 retrieval system \cite{Robertson1976RelevanceWO} that uses sparse features such as word frequencies to calculate similarity with documents. 

\subsubsection{Dense retrieval}
For dense retrieval systems, we fine-tune pre-trained ColBERTv2 \cite{santhanam2021colbertv2} and also use it in the zero-shot setting. Finally, we use the best results for the downstream generation task.

\noindent\textbf{Training:} We fine-tune ColBERTv2 based on triplet formed as $<q, D^+, D^->$. $D^+$ is the document of the libraries relevant to query $q$. $D^-$ is a set of documents of libraries that are not relevant to $q$ but are similar to $D^+$. For $q$ we prepare the training set as ($q$, $d_1^+$, $d_2^+$,.....,$d_m^+$, $d_1^-$, $d_2^-$....., $d_n^-$) where $d_i^+$ is the positive document, and each $d_i^-$ is a negative document which is not relevant to $q$. We select $n$ \emph{hard negatives} using miniLM sentence BERT similarity scores similar to \citet{santhanam2021colbertv2}. Using such a train set, we train ColBERTv2 by minimizing the distance between $q$ and $D^+$ and maximizing the distance between $q$ and $D^-$. 

\subsection{Constrained generation} \label{CD}

Constrained generation is the second stage of \tool{}. It constrains the model during greedy decoding to follow the library grammar using the template, structured schema, and trigger signals. 
In this process, if the model has generated ($x_1$, $x_2$,...$x_n$) tokens, $x_{n+1}$ token is sampled from a set of some specific tokens $t$ such that generated code adheres to the library grammar. This is achieved by setting the logits of all tokens outside $t$ to $-\infty$.

This section explains the steps in constrained generation. First, we explain the \emph{string selection} algorithm, which constrains the model to generate a string from a set of strings. This algorithm will be used repeatedly. Constrained generation starts with fetching the initial templates for $k$ retrieved libraries stored offline. Next, \emph{library selection} algorithm constrains the model to adhere to one of the $k$ library templates. As the model adheres to a template, the \emph{generating variable part} algorithm generates value for the variable part of the template as per the library grammar. While generating the variable part, the guiding template might be changed during decoding based on trigger signals and inter-field dependency as explained by \emph{dynamically changing template} algorithm. Finally, required fields are generated as per \emph{generating required fields} algorithm.


\paragraph{String Selection:} 
\emph{String selection} algorithm is used to constrain the model to generate exactly one string from a set of strings ($S$) \{$s_1, s_2, s_3..., s_n$\} \citep{agrawal2023guiding}. 
Initially, all the strings are tokenized, and we limit the vocabulary $V$ of the model to a set of tokens $t \in V$, which form the prefix of any string in $S$. Once a token $t_i$ among $t$ is sampled, all the strings that do not have $t_i$ as a prefix are discarded. The same process is repeated until exactly one string is chosen.

\paragraph{Library selection:}\label{library selection}We traverse all $k$ initial template strings from left to right and collect substrings for each one until the variable part is encountered. As shown in Figure \ref{Figure 1}, we collect until \emph{gopass}, \emph{lpass}, and \emph{last} as they are static and subsequent parts of text are variable.
As soon as the decoding starts, we constrain the model using \emph{string selection} algorithm to generate exactly one of the $k$ substrings. Next, decoding is constrained to follow that template from left to right while adhering to the grammar of the corresponding library. 

\paragraph{Generating variable part:}Two conditions govern variable part generation. Firstly, when the position and presence of the field are fixed, the model is constrained to select the valid keywords for that using the \emph{string selection} algorithm. 
Secondly, predefined trigger conditions guide the model in generating from specific string pools when the position or presence varies, determined by query $q$. For example, the template of the bash command \emph{gh} is \emph{gh <command> <subcommand> [flags]}. In this example, \emph{<command>}, \emph{<subcommand>}, and \emph{[flags]} are the variable parts. The position and presence of \emph{command} and \emph{subcommand} are fixed, and the model is constrained to select the valid keywords for that part using the string selection algorithm. \emph{Flags} is optional, and a pre-defined trigger condition controls its generation.


\paragraph{Dynamically changing template:}In many cases, one field's presence depends on another. For example, as shown in Figure \ref{Figure 1}, the valid flags and arguments change depending on the sub-command generated. Similarly, in Ansible YAML, the rules of the nested schema (optional and required keys) are completely different from those of the parent schema. Hence, if a key with nested schema is produced, the guiding template is changed to follow the rules of nested schema. After generating each variable part, we check field dependency, and if present, we modify the template accordingly.

\paragraph{Generating required fields:} 
The code must include required fields as per schema rules, but their position is not fixed due to the order-agnostic nature of fields. 
To ensure its presence, we constrain the model to generate the required fields just before the completion of the code. Completion of code is detected by checking for end-of-sequence tokens. This ensures adherence to the schema.

\section{Dataset} \label{Dataset}
This section describes datasets for NL to bash and Ansible YAML task, including augmenting datasets with module descriptions and schema information. 

\subsection{Ansible YAML}

We compile the \nlyaml{} dataset by extracting data from Google BigQuery and Ansible Galaxy. The dataset comprises over $18k$ of NL to YAML samples, sourced from a diverse collection of more than $2500$ modules. We also curate schema rules and descriptions for every module. Schema rules consist of valid optional, required keys and details of the nested schema.
We show dataset statistics in Table \ref{table:data-stats} and more details on data curation in the Appendix \ref{subsec:appendix-ansibleyaml}.

\subsection{Bash command}

Since we primarily focus on improving performance for unseen libraries and low-resource data settings, we select the \tldr{} \citep{zhou2022doccoder} as our primary dataset for NL to Bash. \tldr{} consists of $1503$ bash utilities across the train and test samples. This data consists of $7342$ NL to bash pairs with $4.3$ pairs for every utility. Train and test splits of this data consist of $7342$ NL to bash pairs. A low number of samples for each utility creates a scarce data scenario.

Other than this, we also use \nlbash{} \citep{lin2018nl2bash} dataset consisting of $8090$ train and $609$ test samples for $100$ bash utilities. Due to the high number of NL to bash pairs for every bash utility, this dataset allows us to check performance for resource-rich settings. However, Since this is not the major focus of the work, results for \nlbash{} are included in Appendix (Table \ref{NL2bash})

To prepare module descriptions, we use the \emph{description} section of Linux man-pages \footnote{\url{https://manned.org/pkg/ubuntu-mantic}}.
Further, we augment the \tldr{} dataset with the schema rules for each bash utility. Schema information includes a bash command template prepared from \emph{synopsis} section, valid fields (flags and sub-commands), and inter-field dependency information. Schema details and example templates are provided in \ref{subsec:appendix-nl2bash}.

\begin{table}[!ht]
    \footnotesize 
    \renewcommand{\arraystretch}{1.3} 
	\setlength\tabcolsep{3pt} 
	\begin{tabularx}{\columnwidth}{crrrr} 

	\toprule
        
        \multirow{2}{*}{\thead{\bf Model}}  & \multicolumn{2}{c}{\thead{\bf Bash}} & \multicolumn{2}{c}{\thead{\bf Ansible YAML}}\\
        \cmidrule(l{3mm}r{3mm}){2-3}  \cmidrule(l{3mm}r{3mm}){4-5}
		  & \thead{\bf {Exact}} & \thead{\bf {Token}}  & \thead{\bf {Schema}} & \thead{\bf {Ansible}} \\
           & \thead{\bf Match(\%)} & \thead{\bf F1} & \thead{\bf Correct} &\thead{\bf Aware} \\ 
		\midrule

        GPT Neo 1.3B (*) & 3.23 & 31.97 & 3.11 & 2.51\\
        GPT Neo 1.3B (+) & 4.18 & 32.78 & 4.23 & 3.37\\
        \citet{zhou2022doccoder} & 9.05 & 37.24 & - & -\\
        base+IR & 5.91 & 39.20 & 15.37 & 10.72\\
        base+IR+CD & \textbf{9.40} & \textbf{41.26} & \textbf{36.58} & \textbf{25.19}\\ \hline

        StarCoder2 3B (*) & 4.09 & 34.22 & 4.41 & 5.80 \\
        StarCoder2 3B (+) & 3.38 & 35.53 & 4.96 & 5.90 \\
        base+IR & 7.63 & 41.67 & 7.47 & 4.08\\
        base+IR+CD & \textbf{9.56} & \textbf{43.25} & \textbf{58.82} & \textbf{19.76}\\ \hline
    
        StarCoder2 7B (*) & 4.12 & 34.45 & 5.16 & 5.61 \\
        StarCoder2 7B (+) & 5.49 & 35.72 & 5.11 & 5.63 \\
        base+IR & 8.12 & 42.12 & 22.47 & 11.40\\
        base+IR+CD & \textbf{10.21} & \textbf{44.09} & \textbf{57.00} & \textbf{18.37}\\ \hline
        
        \bottomrule
	\end{tabularx}

\caption{Results for each fine-tuned language model for OOD setting with and without IR and constrained decoding. Here, the model is constrained to follow the Top-1 retrieved library template only. All the metrics in this table demonstrate the syntactic and semantic correctness of the code. \emph{Model (*)} represents the base fine-tuned model and \emph{model (+)} represents the pre-trained fine-tuned model baseline.}

\label{table:OOD}

\end{table}

\section{Experiments}
In this section, we lay out our experiments across \nltocode{} tasks and datasets.

\subsection{Experimental settings}
We evaluate the performance of our framework on two diverse code languages, Ansible-YAML and bash command. For both tasks, we experiment with two settings involving different train-test splits.

\noindent\textbf{Out of Domain:} Here, code libraries in the train and test set are completely disjoint, allowing us to evaluate our method for unseen libraries. We use the original train-test split in \tldr{} dataset for the bash. For YAML, we randomly split the data into $17647$ train and $2056$ test samples with $2483$ libraries in the train and $365$ in the test. OOD split results are demonstrated in Table \ref{table:OOD}.

\noindent\textbf{In Domain:} In this setting, libraries in the test set are a subset of the train set. For bash, we mix the train and test samples of \tldr{} and re-split them in the ratio of 85\% train and 15\% test samples. Further, we filter out the small number of pairs that do not have bash utility in the train set. Finally, we have $6240$ train and $1081$ test NL to bash command pairs with $1503$ unique bash utilities. A similar approach is followed for YAML, which creates $18574$ train and $2989$ test samples. 

\subsection{Baselines}
Across every task and setting, we establish multiple baselines. The Appendix section \ref{subsec:appendix-modeltraininf} describes the hyperparameter details for experiments.


\paragraph{Base (model(*)):} Here, we fine-tune the transformer-based decoder-only model for \nltocode{} tasks.
\paragraph{Base + IR:} We constrain the base fine-tuned model to follow the template of one of the $k$ retrieved libraries as described by the library selection algorithm (refer to \ref{library selection}). However, we do not constrain the model to adhere to its schema for further generation. This allows us to observe the improvement based on the first stage of \tool{} only. Here, we present the results for $k=1$. Results for $k=3,10$ are shown in the Table \ref{OOD top 3}, \ref{Id top 3}. Further details on pre-training data are provided in the Appendix (section \ref{pretrain-ansible}, \ref{pretrain-bash}).
\paragraph{Pre-train (model(+)):} Existing methods like APICoder \citep{zan-etal-2022-language} pre-train models on abundant documentation and code samples for general-purpose languages like Python. Replicating this setup for structured DSLs is challenging due to the scarcity of available code samples. Hence, for best comparison, we pre-train our models on Linux man pages for bash and Ansible documentation for YAML, ensuring no data leakage from fine-tuning datasets. We then fine-tune the pre-trained model on respective \nltocode{} tasks and compare its performance with \tool{}. We also perform ablation studies with Base + IR setup for the pre-trained models (Table \ref{Pre-train OOD top 1}, \ref{Pre-train Id top 1}). Details of pre-training data are provided in the Appendix (section \ref{pretrain-bash}, \ref{pretrain-ansible}).
\paragraph{DocPrompting:} We adopt \docp{} \citep{zhou2022doccoder} as a baseline for OOD split through the TLDR dataset because it is a RAG-based approach, currently state-of-the-art for \tldr{}. Additionally, Unlike other RAG-based methods \citep{parvez2021retrieval, zhang2023repocoder}, it uses documentation instead of abundant code samples, aligning better with our DSL use case with scarce examples.


\begin{table}[!ht]
    \footnotesize 
    \renewcommand{\arraystretch}{1.3} 
	\setlength\tabcolsep{3pt} 
	\begin{tabularx}{\columnwidth}{crrrr} 

	\toprule
        
        \multirow{2}{*}{\thead{\bf Model}}  & \multicolumn{2}{c}{\thead{\bf Bash}} & \multicolumn{2}{c}{\thead{\bf Ansible YAML}}\\
        \cmidrule(l{3mm}r{3mm}){2-3}  \cmidrule(l{3mm}r{3mm}){4-5}
		  & \thead{\bf {Exact}} & \thead{\bf {Token}}  & \thead{\bf {Schema}} & \thead{\bf {Ansible}} \\
           & \thead{\bf Match(\%)} & \thead{\bf F1} & \thead{\bf Correct} &\thead{\bf Aware} \\ 
		\midrule

        GPT Neo 1.3B (*) & 8.08 & 44.02 & 3.11 & 2.51\\
        GPT Neo 1.3B (+) & 9.12 & 45.23 & 4.23 & 3.37\\
        base+IR & 9.12 & 47.13 & 15.37 & 10.72\\
        base+IR+CD & \textbf{10.46} & \textbf{49.37} & \textbf{36.58} & \textbf{25.19}\\ \hline

        StarCoder2 3B (*) & 15.26 & 50.38 & 4.65 & 5.25 \\
        StarCoder2 3B (+) & 15.26 & 51.74 & 4.71 & 6.20 \\
        base+IR & 16.31 & 54.31 & 6.11 & 9.22\\
        base+IR+CD & \textbf{17.23} & \textbf{56.12} & \textbf{51.08} & \textbf{39.04}\\ \hline
    
        StarCoder2 7B (*) & 14.91 & 50.82 & 4.38 & 6.49 \\
        StarCoder2 7B (+) & 15.63 & 52.73 & 4.11 & 6.39 \\
        base+IR & 16.79 & 54.77 & 7.05 & 10.43\\
        base+IR+CD & \textbf{18.12} & \textbf{57.64} & \textbf{52.96} & \textbf{36.94}\\ \hline

        \bottomrule
	\end{tabularx}

\caption{Results for each fine-tuned language model for ID setting with and without IR and constrained decoding. Here, the model is constrained to follow the Top-1 retrieved library template only. All the metrics in this table demonstrate the syntactic and semantic correctness of the code.}

\label{table:In-domain}

\end{table}

\begin{table}[!ht]
    \scriptsize 
    \renewcommand{\arraystretch}{1.2} 
    \setlength\tabcolsep{2pt} 
    \begin{tabularx}{\columnwidth}{crrrr} 

    \toprule
        
        \multirow{2}{*}{\thead{\bf Model}}  & \multicolumn{2}{c}{\thead{\bf OOD}} & \multicolumn{2}{c}{\thead{\bf ID}}\\
        \cmidrule(l{2mm}r{2mm}){2-3}  \cmidrule(l{2mm}r{2mm}){4-5}
          & \thead{\bf {CMD}} & \thead{\bf {Module}}  & \thead{\bf {CMD}} & \thead{\bf {Module}} \\
           & \thead{\bf Acc(\%)} & \thead{\bf Match(\%)} & \thead{\bf Acc(\%)} &\thead{\bf Match(\%)} \\ 
        \midrule

        GPT Neo 1.3B (*) & 17.88 & 18.63 & 37.01 & 32.71\\
        GPT Neo 1.3B (+) & 17.13 & 17.01 & 39.21 & 33.48\\
        StarCoder2 3B (*) & 17.13 & 25.12 & 47.91 & 52.79 \\
        StarCoder2 3B (+) & 17.02 & 26.16 & 48.38 & 53.90\\
        StarCoder2 7B (*) & 16.16 & 22.13 & 46.99 & \textbf{77.95} \\
        StarCoder2 7B (+) & 17.88 & 21.98 & 48.38 & 77.81 \\
        +IR/+IR+CD & \textbf{38.32} & \textbf{36.38} & \textbf{60.12} & 68.45\\

        \bottomrule
    \end{tabularx}

\caption{Results for the library (bash utility or ansible module) detection accuracy in generated code. Here, the model is constrained to follow the Top-1 retrieved library template only. Hence, Command Acc and Module Acc, which detect the exact match of the library in generated code, depend only on IR and give the same scores for IR and IR+CD models.}

\label{table:library-detection}

\end{table}

\subsection{Models}

\textbf{Information Retrieval} We experiment with sparse retrieval BM25 and dense retrieval ColBERTv2. 


\noindent\textbf{Generator} We include different sized state-of-the-art code language models in our evaluation, including StarCoder2 family (3B, 7B, 15B) \citep{lozhkov2024starcoder}, and CodeLlama 34B \citep{roziere2023code}. Due to resource constraints to fine-tune large parameter models like CodeLlama 34B and Starcoder2 15B, we experiment with their instruction-tuned version in a 3-shot setting and present their results in Appendix (Table \ref{OOD in context}). Further, our evaluation includes a fine-tuned GPT Neo 1.3B \citep{gpt-neo} version to compare with the \docp{} baseline. We use beam search inference decoding for all the base fine-tuned models with beam width $5$.

\subsection{Evaluation metrics}
\textbf{IR:} We evaluate IR using Hits@k metric ($k=\{1,3,5\}$). This metric indicates the percentage of accurate documents within the top k retrievals.\\
\textbf{Bash command:} Evaluation metrics for bash include 1) Command name accuracy (CMD Acc): This metric evaluates the exact match of bash utility in the command (e.g. \emph{tar, cat}). 2) Exact Match: Exact match of full generated command and reference command 3) Token F1 score \citep{zhou2022doccoder}.

\noindent\textbf{Ansible YAML:} We leverage $2$ evaluation metrics from \citet{10247987} - \schemacorrect{}, and \ansbileaware{}. Additionally, we introduce the \modulematch{} metric, which measures the correctness of the generated YAML module. This metric is similar to the CMD Acc metric in bash. Refer to \ref{eval metric} for a detailed description of metrics.

\section{Results and Analysis}
Results and comparison of our framework with various baselines are presented in Tables \ref{table:OOD}, \ref{table:In-domain} and \ref{table:library-detection}. This section presents several observations and a qualitative analysis of the performance.

\paragraph{Improvement in module accuracy:} We observe that extended pre-training does not improve performance in structured DSLs with limited code samples in the documentation. Therefore, we use an IR-based approach that focuses on retrieving utility descriptions, unlike \citet{zhou2022doccoder}, which retrieves passages with options (flags and sub-commands) and utilities. This targeted detection reduces the search space for IR from $400k$ to $1.5k$ documents, leading to a notable improvement in Hits@1 (Table \ref{table:ir-results}). This improves CMD Acc from $27.59\%$ to $38.32\%$ when the model is constrained to follow the Hits@1 retrieved library template (Table \ref{table:library-detection}). CMD Acc consistently improves for the ID setting by around $6\%$ to $12\%$ (Table \ref{table:library-detection}). For YAML, \modulematch{} significantly improves compared to the fine-tuned baselines, especially in the OOD setting ($\sim10\%$). Further, we restrict the model to follow one of the templates for $k$ retrieved libraries. CMD Acc and \modulematch{} drop with a higher value of $k$ (Table \ref{OOD top 3}, \ref{Id top 3}), which is expected since relaxing constraints on the model tend to approach its performance towards the baselines.

\begin{table*}[!ht]
\centering
\begin{tabular}{clcllcl|lcllcl}
\toprule
 &
  \multicolumn{6}{c|}{\small \textbf{Bash}} &
  \multicolumn{6}{c}{\small \textbf{Ansible YAML}} \\ \cline{2-13} 
 &
   \multicolumn{6}{c|}{\small \textbf{Hits@k}} &
  \multicolumn{6}{c}{\small \textbf{Hits@k}} \\ 
  \cline{2-13} &
  \multicolumn{3}{c}{\small \textbf{In Domain}} &
  \multicolumn{3}{c|}{\small \textbf{Out of Domain}} &
  \multicolumn{3}{c}{\small \textbf{In Domain}} &
  \multicolumn{3}{c}{\small \textbf{Out of Domain}} \\ \cline{2-13} 
\multicolumn{1}{l}{} &
  \scriptsize @1 &
  \multicolumn{1}{l}{\scriptsize @3} &
  \scriptsize @10 &
  \scriptsize @1 &
  \multicolumn{1}{l}{\scriptsize @3} &
  \scriptsize @10 &
  \scriptsize @1 &
  \multicolumn{1}{l}{\scriptsize @3} &
  \scriptsize @10 &
  \scriptsize @1 &
  \multicolumn{1}{l}{\scriptsize @3} &
  \scriptsize @10 \\ \hline
\small BM25 &
  \small 43.21 &
  \small 56.78 &
  \small 68.34 &
  \small 14.51 &
  \small 21.65 &
  \small 32.57 &
  \small 20.51 &
  \small 30.11 &
  \small 39.78 &
  \small 16.20 &
  \small 24.37 &
  \small 33.12 \\ \hline
\begin{tabular}[c]{@{}c@{}}\small ColBERTv2\\ \small (Zero Shot)\end{tabular} &
  \small 53.43 &
  \small 71.26 &
  \small 78.90 &
  \small \textbf{38.32} &
  \small \textbf{51.78} &
  \small \textbf{58.76} &
  \small 37.69 &
  \small 50.24 &
  \small 61.99 &
  \small 30.30 &
  \small 42.31 &
  \small 55.65 \\ \hline
  \begin{tabular}[c]{@{}c@{}}\small ColBERTv2\\ \small (Fine-tuned)\end{tabular}&
  \small \textbf{61.62} &
  \small \textbf{79.23} &
  \small \textbf{84.56} &
  \small 32.21 &
  \small 47.81 &
  \small 54.28 &
  \textbf{\small 66.54} &
  \textbf{\small 77.42} &
  \textbf{\small 84.81} &
  \textbf{\small 34.58} &
  \textbf{\small 47.61} &
  \multicolumn{1}{c}{\textbf{\small 58.46}} \\ \bottomrule
\end{tabular}
\caption{Performance of sparse and dense retrieval across \nltocode{} tasks for ID and OOD settings.}

\label{table:ir-results}
\end{table*}
\paragraph{Improvement in Code:} 

In the OOD setting (Table \ref{table:OOD}), fine-tuned code LM baselines struggle to generate correct libraries even for popular languages like Bash, eventually leading to semantically poor code not relevant to the NL query. While, in the ID setting, despite generating correct libraries (indicated by high \modulematch{} or CMD Acc), baseline models struggle to generate syntactically correct intended code, resulting in subpar Token F1, \schemacorrect{}, and \ansbileaware{} metric scores (Table \ref{table:In-domain}). This is more pronounced in YAML due to its complex format and diverse schemas.
Constraining the model to follow schema rules during decoding restricts the generation of invalid keywords and significantly improves performance across all metrics and settings.
For bash, we observe significant improvement (Table \ref{table:OOD}) over \docp{} in Token F1 score by leveraging grammar templates from the documentation. For example, for the NL query, \emph{reboot the device from fastboot mode into fastboot mode again}, the ground truth command is shown in Listing \ref{lst:fastbootexample}.

\begin{lstlisting}[language=bash, caption=Example sample for fastboot command, label={lst:fastbootexample}]
# ground truth command
fastboot reboot bootloader

# DocPrompting output command
fastboot reboot path/to/devicefile

# example fastboot command template
fastboot [flags] <flashall|erase partition|flashing unlock|reboot bootloader|...>

# DocCGen output command
fastboot reboot bootloader
\end{lstlisting}
\docp{} retrieves correct documents for the given query, which consists of the description of the utility \emph{fastboot} and a document for the subcommand fields \emph{reboot}. Yet it produces an incorrect command as shown in the Listing \ref{lst:fastbootexample}. We instead leverage the template from the \emph{synopsis} and \emph{commands} section of fastboot documentation. As shown in Listing \ref{lst:fastbootexample}, following the grammar template ensures that subcommand is generated from valid strings enclosed in <>. This ensures \emph{reboot} is followed by the word \emph{bootloader}. This approach improves the Token F1 score from $37.24$ to $41.26$. Hence, constrained decoding using the templates and schema rules reduces the generation of invalid keywords resulting in improved validity of code and agreement with ground truth.


\section{Conclusion}

We propose \tool{}, a novel framework for \nltocode{} generation for structured DSLs. \tool{} decomposes the \nltocode{} generation into two steps involving the detection of relevant libraries in the first step and using schema and grammar rules extracted from the documentation of these libraries to guide the decoding in the second step.
We evaluate the performance of \tool{} for two complex structured languages, Bash command and Ansible YAML, involving two settings, OOD and ID. Our approach outperforms state-of-the-art techniques consistently across all metrics for different-sized models. It reduces syntactic and semantic errors in code, particularly for unseen libraries and low-resource data settings. We also contribute the first \emph{publicly} available benchmark dataset for NL to Ansible-YAML task. We augment NL to Ansible-YAML and \tldr{} dataset with description and schema information. We hope this work will help advance research in solving DSL-related tasks and constrained generation.

\section*{Limitations}
We break down code generation in to two steps: a) Information Retrieval and b) Generation based on retrieved documentation. Therefore, errors in retrieval for the user query may cascade to the generation step. Even though, we see that leveraging documentation in this pipeline-based approach results in significant improvements for custom settings, we believe that jointly training the retriever and generator might mitigate these errors. This can be explored as a part of future work. Apart from this, constrained decoding adds a computational overhead during inference. However, since we add the rules on top of efficient greedy decoding, constrained decoding is practical to use as beam search decoding which is widely adopted is similarly computationally heavy. Still, this can be mitigated using constrained generation in speculative decoding similar to \citet{wang2024grammar}. Such improvements can easily be integrated with our framework. Further, parser-based methods to automatically integrate grammar rules during decoding can help generalize \tool{} to a larger scale.

\section*{Ethics Statement}
Custom curated \nlyaml{} data has been collected from sources like Google BigQuery and Ansible Galaxy, which are publicly available platforms. Other datasets and documents used are from open-source repositories, are publicly available, and can be used without any copyright issues.
\bibliography{anthology,custom}

\begin{thebibliography}{19}
\expandafter\ifx\csname natexlab\endcsname\relax\def\natexlab#1{#1}\fi

\bibitem[{Agrawal et~al.(2023)Agrawal, Kanade, Goyal, Lahiri, and Rajamani}]{agrawal2023guiding}
Lakshya~A Agrawal, Aditya Kanade, Navin Goyal, Shuvendu~K Lahiri, and Sriram~K Rajamani. 2023.
\newblock Guiding language models of code with global context using monitors.
\newblock \emph{arXiv preprint arXiv:2306.10763}.

\bibitem[{Bhaskar et~al.(2023)Bhaskar, Tomar, Sathe, and Sarawagi}]{bhaskar2023benchmarking}
Adithya Bhaskar, Tushar Tomar, Ashutosh Sathe, and Sunita Sarawagi. 2023.
\newblock Benchmarking and improving text-to-sql generation under ambiguity.
\newblock \emph{arXiv preprint arXiv:2310.13659}.

\bibitem[{Black et~al.(2021)Black, Leo, Wang, Leahy, and Biderman}]{gpt-neo}
Sid Black, Gao Leo, Phil Wang, Connor Leahy, and Stella Biderman. 2021.
\newblock \href {https://doi.org/10.5281/zenodo.5297715} {{GPT-Neo: Large Scale Autoregressive Language Modeling with Mesh-Tensorflow}}.
\newblock {If you use this software, please cite it using these metadata.}

\bibitem[{Ding et~al.(2022)Ding, Wang, Ahmad, Ramanathan, Nallapati, Bhatia, Roth, and Xiang}]{ding2022cocomic}
Yangruibo Ding, Zijian Wang, Wasi~Uddin Ahmad, Murali~Krishna Ramanathan, Ramesh Nallapati, Parminder Bhatia, Dan Roth, and Bing Xiang. 2022.
\newblock Cocomic: Code completion by jointly modeling in-file and cross-file context.
\newblock \emph{arXiv preprint arXiv:2212.10007}.

\bibitem[{Lin et~al.(2018)Lin, Wang, Zettlemoyer, and Ernst}]{lin2018nl2bash}
Xi~Victoria Lin, Chenglong Wang, Luke Zettlemoyer, and Michael~D Ernst. 2018.
\newblock Nl2bash: A corpus and semantic parser for natural language interface to the linux operating system.
\newblock \emph{arXiv preprint arXiv:1802.08979}.

\bibitem[{Lozhkov et~al.(2024)Lozhkov, Li, Allal, Cassano, Lamy-Poirier, Tazi, Tang, Pykhtar, Liu, Wei et~al.}]{lozhkov2024starcoder}
Anton Lozhkov, Raymond Li, Loubna~Ben Allal, Federico Cassano, Joel Lamy-Poirier, Nouamane Tazi, Ao~Tang, Dmytro Pykhtar, Jiawei Liu, Yuxiang Wei, et~al. 2024.
\newblock Starcoder 2 and the stack v2: The next generation.
\newblock \emph{arXiv preprint arXiv:2402.19173}.

\bibitem[{Lu et~al.(2022)Lu, Duan, Han, Guo, Hwang, and Svyatkovskiy}]{lu2022reacc}
Shuai Lu, Nan Duan, Hojae Han, Daya Guo, Seung-won Hwang, and Alexey Svyatkovskiy. 2022.
\newblock Reacc: A retrieval-augmented code completion framework.
\newblock \emph{arXiv preprint arXiv:2203.07722}.

\bibitem[{Parvez et~al.(2021)Parvez, Ahmad, Chakraborty, Ray, and Chang}]{parvez2021retrieval}
Md~Rizwan Parvez, Wasi~Uddin Ahmad, Saikat Chakraborty, Baishakhi Ray, and Kai-Wei Chang. 2021.
\newblock Retrieval augmented code generation and summarization.
\newblock \emph{arXiv preprint arXiv:2108.11601}.

\bibitem[{Poesia et~al.(2022)Poesia, Polozov, Le, Tiwari, Soares, Meek, and Gulwani}]{poesia2022synchromesh}
Gabriel Poesia, Oleksandr Polozov, Vu~Le, Ashish Tiwari, Gustavo Soares, Christopher Meek, and Sumit Gulwani. 2022.
\newblock Synchromesh: Reliable code generation from pre-trained language models.
\newblock \emph{arXiv preprint arXiv:2201.11227}.

\bibitem[{Pujar et~al.(2023)Pujar, Buratti, Guo, Dupuis, Lewis, Suneja, Sood, Nalawade, Jones, Morari, and Puri}]{10247987}
Saurabh Pujar, Luca Buratti, Xiaojie Guo, Nicolas Dupuis, Burn Lewis, Sahil Suneja, Atin Sood, Ganesh Nalawade, Matt Jones, Alessandro Morari, and Ruchir Puri. 2023.
\newblock \href {https://doi.org/10.1109/DAC56929.2023.10247987} {Invited: Automated code generation for information technology tasks in yaml through large language models}.
\newblock In \emph{2023 60th ACM/IEEE Design Automation Conference (DAC)}, pages 1--4.

\bibitem[{Robertson and Jones(1976)}]{Robertson1976RelevanceWO}
Stephen~E. Robertson and Karen~Sp{\"a}rck Jones. 1976.
\newblock \href {https://api.semanticscholar.org/CorpusID:45186038} {Relevance weighting of search terms}.
\newblock \emph{J. Am. Soc. Inf. Sci.}, 27:129--146.

\bibitem[{Roziere et~al.(2023)Roziere, Gehring, Gloeckle, Sootla, Gat, Tan, Adi, Liu, Remez, Rapin et~al.}]{roziere2023code}
Baptiste Roziere, Jonas Gehring, Fabian Gloeckle, Sten Sootla, Itai Gat, Xiaoqing~Ellen Tan, Yossi Adi, Jingyu Liu, Tal Remez, J{\'e}r{\'e}my Rapin, et~al. 2023.
\newblock Code llama: Open foundation models for code.
\newblock \emph{arXiv preprint arXiv:2308.12950}.

\bibitem[{Santhanam et~al.(2021)Santhanam, Khattab, Saad-Falcon, Potts, and Zaharia}]{santhanam2021colbertv2}
Keshav Santhanam, Omar Khattab, Jon Saad-Falcon, Christopher Potts, and Matei Zaharia. 2021.
\newblock Colbertv2: Effective and efficient retrieval via lightweight late interaction.
\newblock \emph{arXiv preprint arXiv:2112.01488}.

\bibitem[{Scholak et~al.(2021)Scholak, Schucher, and Bahdanau}]{scholak-etal-2021-picard}
Torsten Scholak, Nathan Schucher, and Dzmitry Bahdanau. 2021.
\newblock \href {https://doi.org/10.18653/v1/2021.emnlp-main.779} {{PICARD}: Parsing incrementally for constrained auto-regressive decoding from language models}.
\newblock In \emph{Proceedings of the 2021 Conference on Empirical Methods in Natural Language Processing}, pages 9895--9901, Online and Punta Cana, Dominican Republic. Association for Computational Linguistics.

\bibitem[{Wang et~al.(2024)Wang, Wang, Wang, Cao, A~Saurous, and Kim}]{wang2024grammar}
Bailin Wang, Zi~Wang, Xuezhi Wang, Yuan Cao, Rif A~Saurous, and Yoon Kim. 2024.
\newblock Grammar prompting for domain-specific language generation with large language models.
\newblock \emph{Advances in Neural Information Processing Systems}, 36.

\bibitem[{Wolf et~al.(2020)Wolf, Debut, Sanh, Chaumond, Delangue, Moi, Cistac, Rault, Louf, Funtowicz, Davison, Shleifer, von Platen, Ma, Jernite, Plu, Xu, Le~Scao, Gugger, Drame, Lhoest, and Rush}]{wolf-etal-2020-transformers}
Thomas Wolf, Lysandre Debut, Victor Sanh, Julien Chaumond, Clement Delangue, Anthony Moi, Pierric Cistac, Tim Rault, Remi Louf, Morgan Funtowicz, Joe Davison, Sam Shleifer, Patrick von Platen, Clara Ma, Yacine Jernite, Julien Plu, Canwen Xu, Teven Le~Scao, Sylvain Gugger, Mariama Drame, Quentin Lhoest, and Alexander Rush. 2020.
\newblock \href {https://doi.org/10.18653/v1/2020.emnlp-demos.6} {Transformers: State-of-the-art natural language processing}.
\newblock In \emph{Proceedings of the 2020 Conference on Empirical Methods in Natural Language Processing: System Demonstrations}, pages 38--45, Online. Association for Computational Linguistics.

\bibitem[{Zan et~al.(2022)Zan, Chen, Lin, Guan, Yongji, and Lou}]{zan-etal-2022-language}
Daoguang Zan, Bei Chen, Zeqi Lin, Bei Guan, Wang Yongji, and Jian-Guang Lou. 2022.
\newblock \href {https://doi.org/10.18653/v1/2022.findings-emnlp.21} {When language model meets private library}.
\newblock In \emph{Findings of the Association for Computational Linguistics: EMNLP 2022}, pages 277--288, Abu Dhabi, United Arab Emirates. Association for Computational Linguistics.

\bibitem[{Zhang et~al.(2023)Zhang, Chen, Zhang, Liu, Zan, Mao, Lou, and Chen}]{zhang2023repocoder}
Fengji Zhang, Bei Chen, Yue Zhang, Jin Liu, Daoguang Zan, Yi~Mao, Jian-Guang Lou, and Weizhu Chen. 2023.
\newblock Repocoder: Repository-level code completion through iterative retrieval and generation.
\newblock \emph{arXiv preprint arXiv:2303.12570}.

\bibitem[{Zhou et~al.(2022)Zhou, Alon, Xu, JIang, and Neubig}]{zhou2022doccoder}
Shuyan Zhou, Uri Alon, Frank~F Xu, Zhengbao JIang, and Graham Neubig. 2022.
\newblock Doccoder: Generating code by retrieving and reading docs.
\newblock \emph{arXiv preprint arXiv:2207.05987}.

\end{thebibliography}
\bibliographystyle{acl_natbib}

\appendix

\section{Appendix}
We provide additional details for \nlyaml{}, and NL to Bash task, hyper-parameter details, and additional analysis on performance in a low resource setting. Firstly we present the details of Ansible-YAML which consists of data collection, schema rules, a list of trigger signals, and evaluation metrics in section \ref{ansibleyaml}. We present the same details for the NL to Bash task in the section \ref{nl2bash}. The appendix also consists of results for additional ablation studies like Top-3, Top-10 IR (Table \ref{OOD top 3}, \ref{Id top 3}) results of in-context learning (Table \ref{OOD in context}), and ablation studies with pre-training data (Table \ref{Pre-train OOD top 1}, \ref{Pre-train Id top 1}).

\label{sec:appendix}

\subsection{Ansible YAML}\label{ansibleyaml}
\label{subsec:appendix-ansibleyaml}

YAML is one of the standard code languages used to configure systems declaratively. Ansible is an IT automation tool widely used in enterprises that allows the Infrastructure as Code (IaC) paradigm through Ansible playbooks written in YAML. This section describes examples, data collection, statistics, and evaluation metrics for \nlyaml{} task.

\subsubsection{Examples}
Some examples (Listing \ref{lst:yamlexample1} and \ref{lst:yamlexample2}) of Ansible YAML are provided to show glimpse of their syntax.

\begin{lstlisting}[caption=Example Ansible YAML for file module with simple key value pairs, label={lst:yamlexample1}]
- name: Create a symbolic link
  ansible.builtin.file:
    src: /file/to/link/to
    dest: /path/to/symlink
    owner: foo
    group: foo
    state: link
\end{lstlisting}

\begin{lstlisting}[caption=Example Ansible YAML for make module with nested key value pairs, label={lst:yamlexample2}]
- name: Build 'all' target with args
  make:
    chdir: /home/ubuntu/cool-project
    target: all
    params:
      NUM_THREADS: 4
      BACKEND: lapack
\end{lstlisting}

\subsubsection{Data Collection}
We curate the dataset from $2$ different sources - Google BigQuery and Ansible Galaxy. To curate data from Google BigQuery, we run a SQL query against the BigQuery datastore to pull code files with one of the valid YAML file extensions (.yaml, .yml, .YAML, and .YML). There is no foolproof way to identify Ansible-YAMLs from this corpus. Therefore, we employ simple heuristics based on module keywords and the format of the data to extract Ansible-YAML candidates. 

From each Ansible YAML file to subsample NL to YAML candidates, we use a heuristic based on YAMLs having the keys - \emph{name} and \emph{name of the ansible module}. These candidates are then grouped based on the ansible module name and then used for preparing in and out-of-domain settings.

A universal set of Ansible modules is fetched from Ansible Galaxy API along with their documentation. The documentation consists of long and short descriptions, module constraints, and examples. The long and short descriptions are used to prepare data for IR. Examples are combined into \nlyaml{} dataset prepared using Google BigQuery, and module constraints are used in the constrained generation stage.

\subsubsection{Data Statistics}
\begin{table}[ht]
    \begin{tabular}{ccccc}
    	\toprule

                                                                         & \multicolumn{2}{c}{\small \bf{In Domain}} &  \multicolumn{2}{c}{\small \bf{Out of Domain}} \\ \cline{2-5} 
                                                                         & \small \bf{Train}          & \small \bf{Test}         & \small \bf{Train}            & \bf{\small Test}           \\ \hline
\small No. of modules                                                           & \small 2922           & \small 2097         & \small 2483             & \small 365            \\ \hline
\small No. of samples                                                           & \small 18574          & \small 2989         & \small 17647            & \small 2056           \\ \hline
\begin{tabular}[c]{@{}c@{}}\small Min no. of samples \\ \small per module\end{tabular} & \small 4              & \small 1            & \small 4                & \small 1              \\ \hline
\begin{tabular}[c]{@{}c@{}}\small Max no. of samples \\ \small per module\end{tabular} & \small 7              & \small 7            & \small 8                & \small 8              \\ \hline
\begin{tabular}[c]{@{}c@{}}\small Average no. of \\ \small samples per module\end{tabular} & \small 6    & \small 1  & \small 7   & \small 6   \\ \hline
\begin{tabular}[c]{@{}c@{}}\small Min no. of \\ \small key value pairs\end{tabular}    & \small 0              & \small 1            & \small 0                & \small 1              \\ \hline
\begin{tabular}[c]{@{}c@{}}\small Max no. of \\ \small key values pairs\end{tabular}       & \small 1225 & \small 97 & \small 187 & \small 111 \\ \hline
\begin{tabular}[c]{@{}c@{}}\small Average no. of \\ \small key value pairs\end{tabular}    & \small 4    & \small 5  & \small 4   & \small 5   \\ 	\bottomrule
\end{tabular}
\label{YAML data Statistics}
\caption{Statistics for \nlyaml{} dataset.}

\label{table:data-stats}
\end{table}
Ansible module, \nlyaml{} sample, and YAML key-value pair distribution are shown in Table \ref{table:data-stats} for both in and out-of-domain settings. The number of samples per module in both settings does not exceed $8$, portraying a low-resource environment. 

Some samples have $0$ key-value pairs because they are simple strings that still are valid YAMLs. The reason for the total number of modules not being consistent across in-domain and out-of-domain settings is that in the out-of-domain setting for test split, some modules have been dropped as the YAMLs were not valid, and similar data processing has been applied to the in-domain setting as well. Also, the number of modules across the splits for the in-domain setting is not equal because the modules having just $1$ sample have been moved to train split to hold the nature of the in-domain setting for the dataset.  

\subsubsection{Module Description and Structured schema}
Ansible Galaxy's API exposes a list of modules and their respective documentation. We use the API to fetch a complete list of modules, and then, for each module, we fetch the module documentation, which includes long and short descriptions. We prepare the module description by appending the short description followed by the long description. We omit those modules which have neither relevant short nor long descriptions. The average length of text descriptions is $816$ characters.

We curate schema information from Ansible Galaxy's API, which returns this information as part of the documentation. We augment the dataset with this schema information, which can include valid required and optional keys as shown in Listing \ref{lst:simpleconst} and nested schema as shown in Listing \ref{lst:nestedconst}. Every nested schema further consists of optional and required keys.

\begin{lstlisting}[caption=Example of type and required key constraints for module device\_administration\_authentication\_rules, label={lst:simpleconst}]
...
"ise_hostname": {
    "description": [
        "The Identity Services Engine hostname."
    ],
    "required": true,
    "type": "str"
},
...
\end{lstlisting}

\begin{lstlisting}[caption=Example of nested key constraints for module device\_administration\_authentication\_rules, label={lst:nestedconst}]
...
"link": {
    "description": "Device Administration Authentication Rules's link.",
    "suboptions": {
        "href": {
            "description": "Device Administration Authentication Rules's href.",
            "type": "str"
        },
        "rel": {
            "description": "Device Administration Authentication Rules's rel.",
            "type": "str"
        },
        "type": {
            "description": "Device Administration Authentication Rules's type.",
            "type": "str"
        }
    },
    "type": "dict"
},
...
\end{lstlisting}

\begin{lstlisting}[caption=Example prompts for \nlyaml{} task, label={lst:nlyamlpromptdescription}]
# array type
- name: Create a symbolic link
...

# dictionary type
name: Create a symbolic link
...
\end{lstlisting}

\paragraph{Trigger signals: }\label{YAML triggers}Trigger signals $G$ for YAML are as follows. If the model produces indentation spaces equal to level one keys, it triggers to constrain the model to produce a valid level one schema by generating valid level 1 keys. Further, if the model generates more spaces, we check the rules for nested schema and constrain the model to adhere to it. If the model generates an invalid indentation, we backtrack, clear the cache of the model, and add the appropriate number of closest indentations in the output. The process of triggering schema rules based on indentation starts to repeat after it.

\begin{lstlisting}[caption=Example Ansible YAML for file module with simple key value pairs. Here{,} {[a|b|c]} denotes one of the values among {a,b,c} is generated. \emph{gen arg} denotes the argument generated without constraints. The key-value pairs for the next line are controlled again based on indentation generated at the end of the argument., label={lst:yamltemplate1}]
- name: Create a symbolic link
  ansible.builtin.file:
    [force|src|dest|owner|group|state....]: {{gen arg}}

- name: Build 'all' target with args
  make:
    [file|chdir|jobs|make|params|target|targets]: {{gen arg}}
\end{lstlisting}

\paragraph{Enforced schema rules: } We ensure that keys generated at every level of YAML adhere to the module schema. YAML consists of optional and required keys. Hence, we ensure that the required keys must be generated in the YAML. We also ensure that none of the keys are duplicated at any level of nesting. The scenario of optional and required keys is followed in the nested schema with keys different than the parent keys. Hence, we follow the rules of nested schema at every level.

\subsubsection{Prompt Description}

In the case of \nlyaml{} task, the prompt is essentially a key-value pair in the YAML, where the key is $name$ and the value is the NL query. The YAML can be an array with one dictionary or a dictionary itself. We show an example in the Listing \ref{lst:nlyamlpromptdescription}.

\subsubsection{Evaluation Metrics}\label{eval metric}

\schemacorrect{} metric evaluates the model on generating schema-compliant YAML, reflecting the YAML's acceptability by the Ansible tool. The \ansbileaware{} metric captures the closeness of the generated YAML to the ground truth by capturing the coverage of the keys and values in the ground truth. We have not used the Exact Match metric from the original paper as it does not capture the nature of Ansible module keys, which are typically order agnostic. We introduce \modulematch{} metric, which evaluates the model's capability to generate the expected module for the given prompt.

\subsection{Pre-training data}\label{pretrain-ansible}

For ansible pre-training, we append the schema information and descriptions for $2.5k$ modules in a text file \footnote{\url{https://docs.ansible.com/ansible/2.9/modules/list_of_all_modules.html}}. We separate the description and schema information in one document by a newline character and two different ansible documents by two newline characters. We observe that this helps the model better learn the domain knowledge. From every documentation we filter code examples as most of the code examples in the Ansible playbook are present in our custom-curated dataset which we use for fine-tuning. The final pre-training dataset consists of $4.14$ million tokens.

\begin{figure*}
\begin{tabular}{ccc}
  \includegraphics[width=50mm]{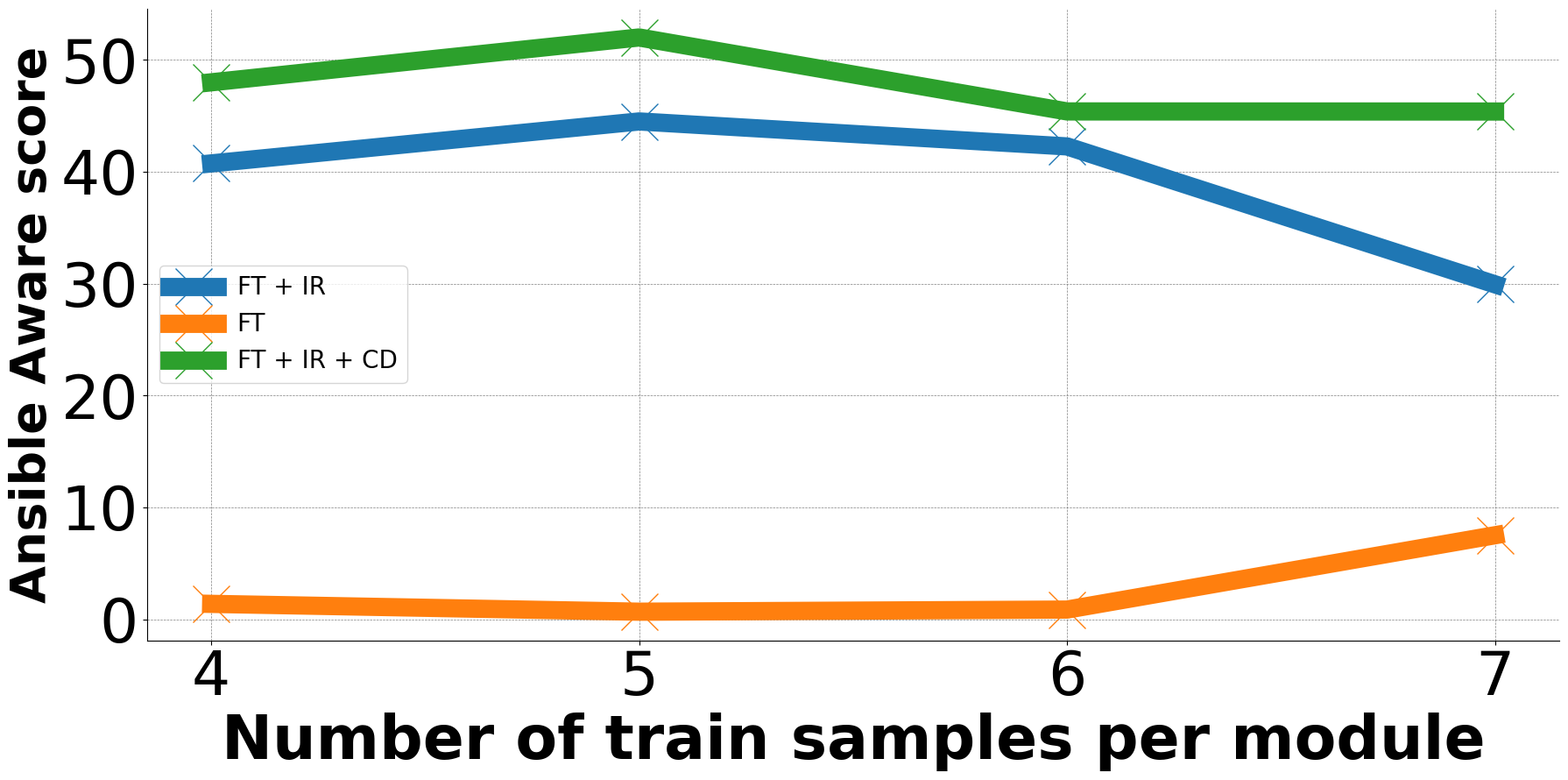} &   \includegraphics[width=50mm]{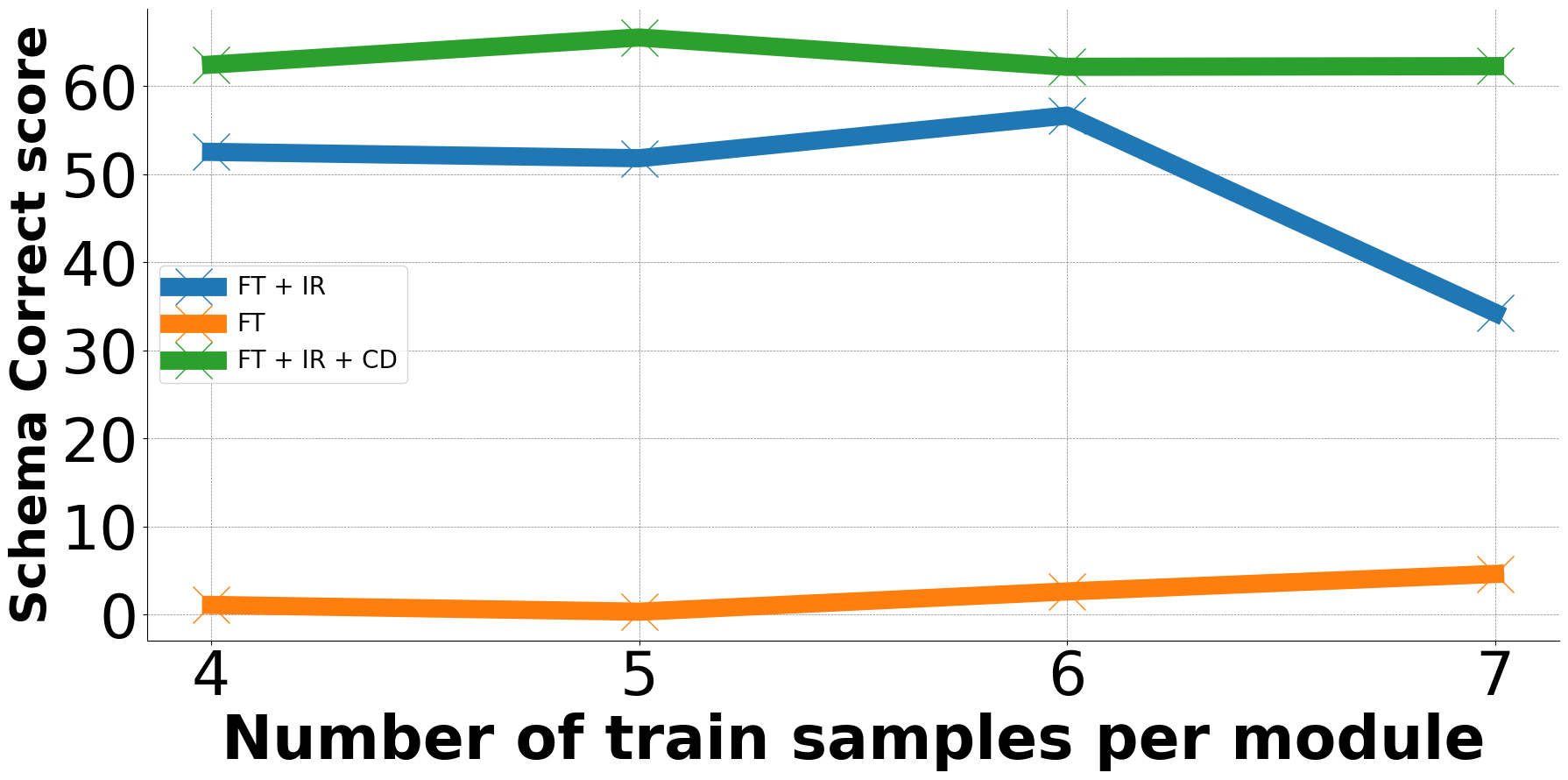} & \includegraphics[width=50mm]{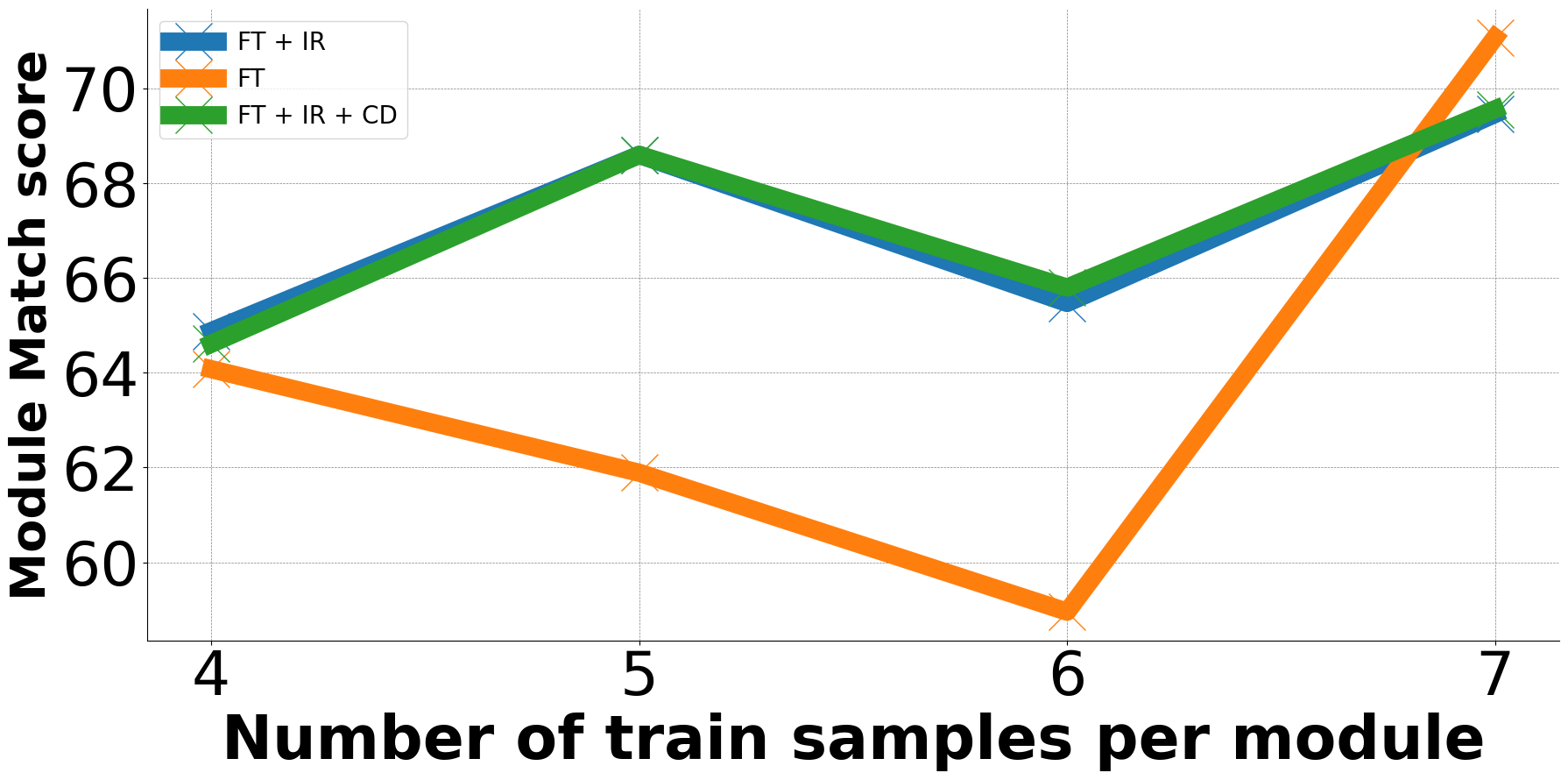}  \\
(a) & (b) & (c) \\
\end{tabular}
\caption{Demonstration of the performance of StarCoder 1B for \nlyaml{} task over varying number of train samples per module for in domain setting.}
\label{plot:lowresource}
\end{figure*}

\begin{figure*}[!ht]
\begin{tabular}{ccc}
  \includegraphics[width=50mm]{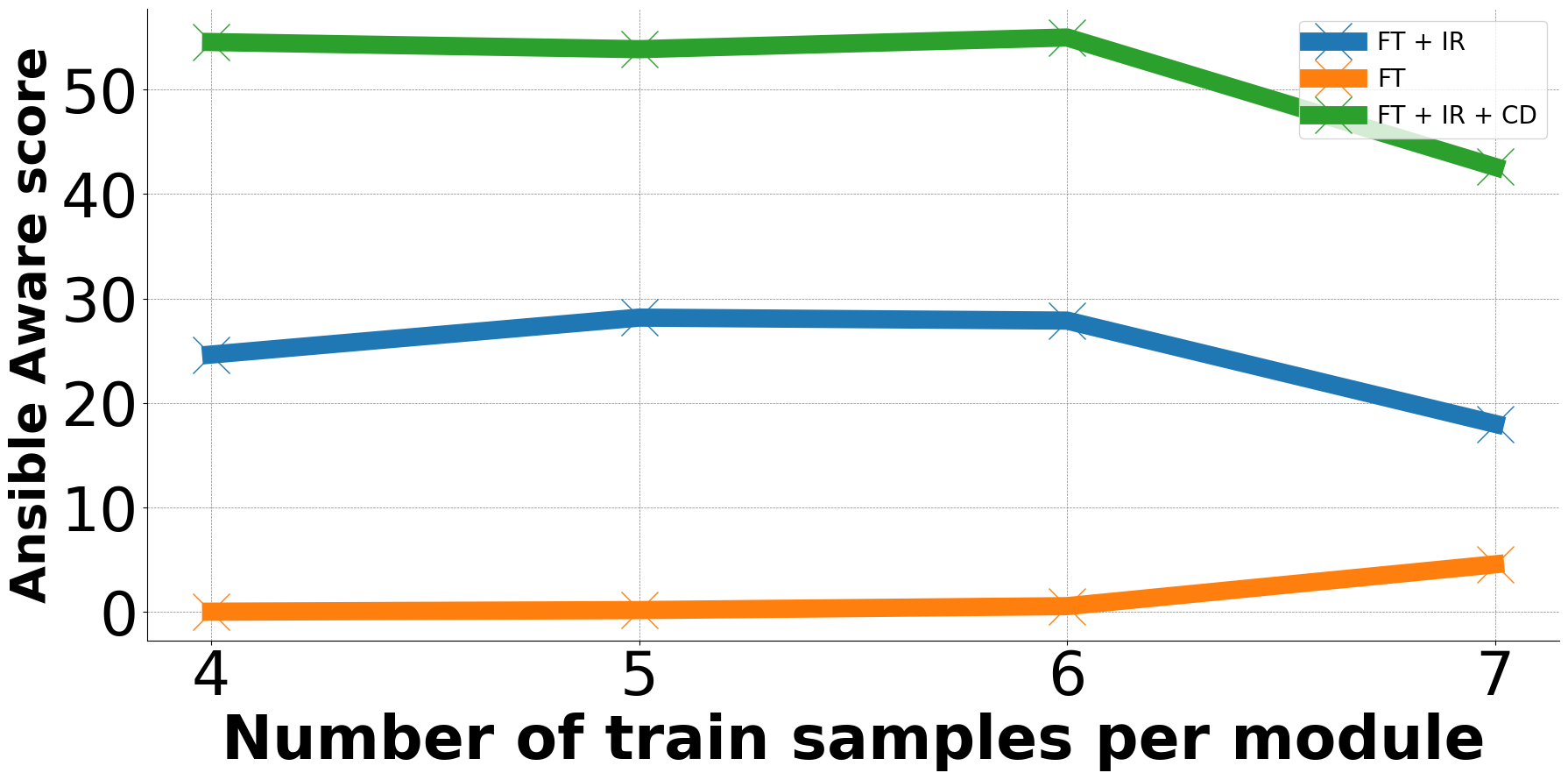} &   \includegraphics[width=50mm]{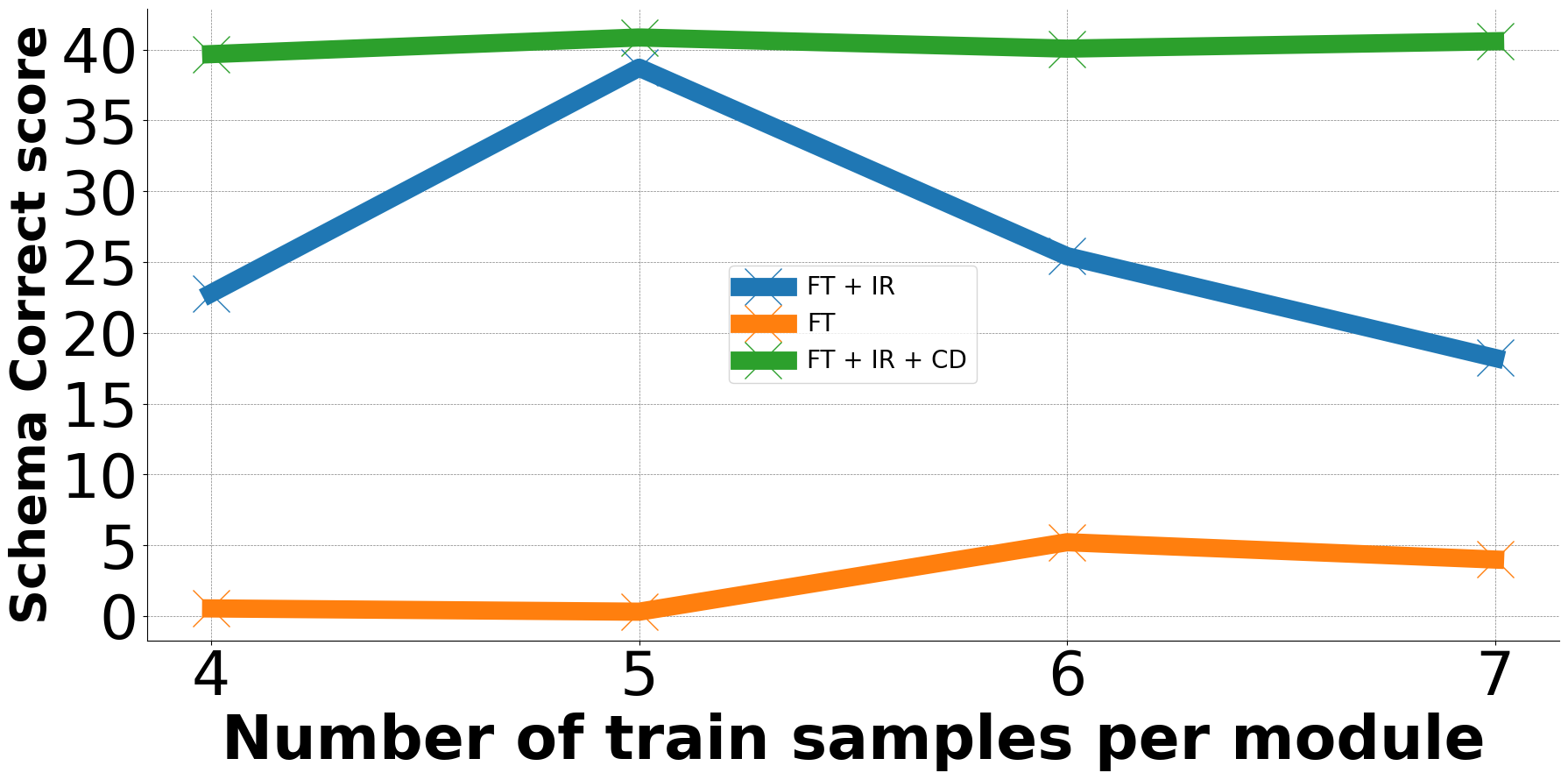} & \includegraphics[width=50mm]{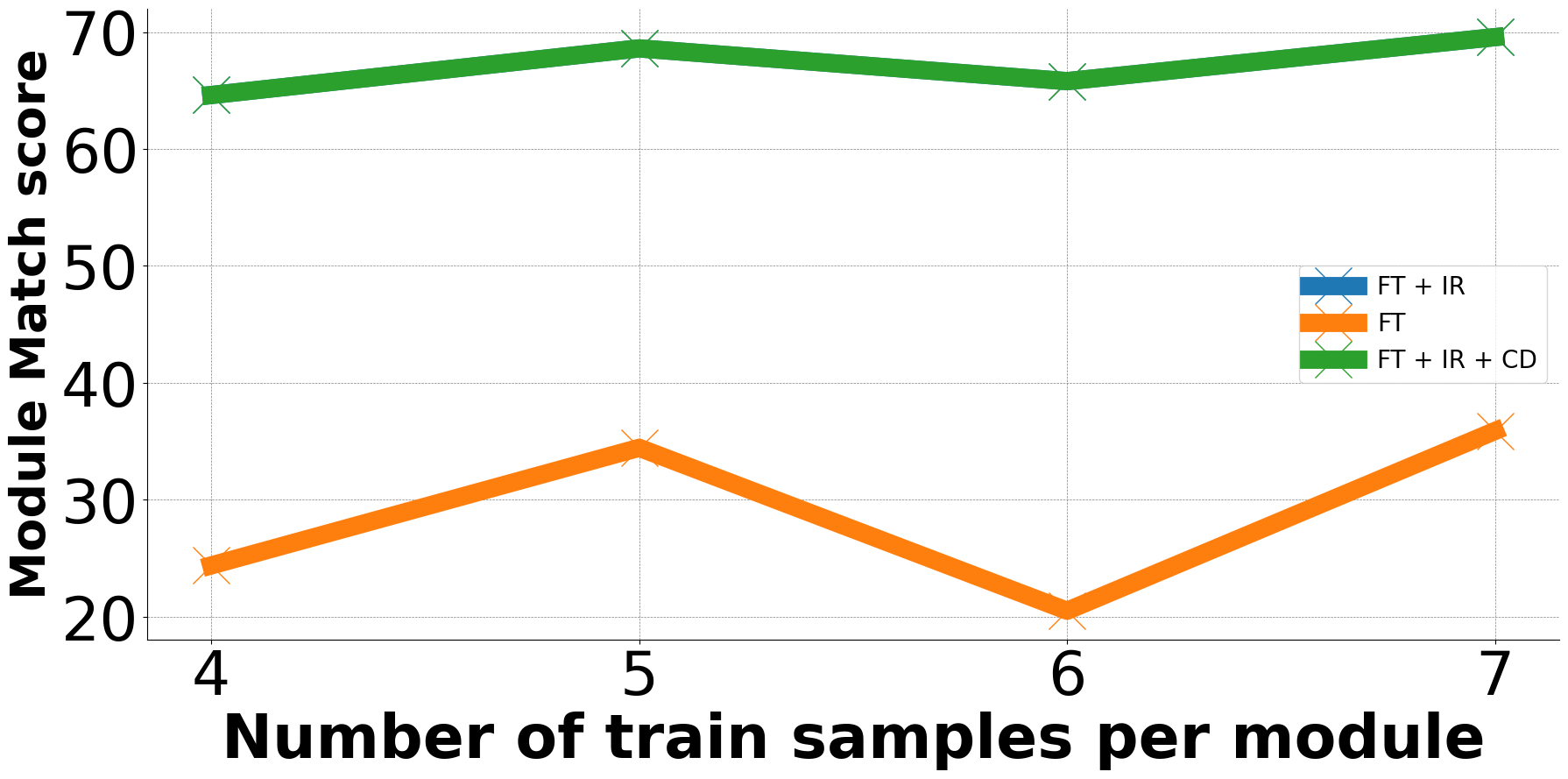}  \\
(a) & (b) & (c) \\
  \includegraphics[width=50mm]{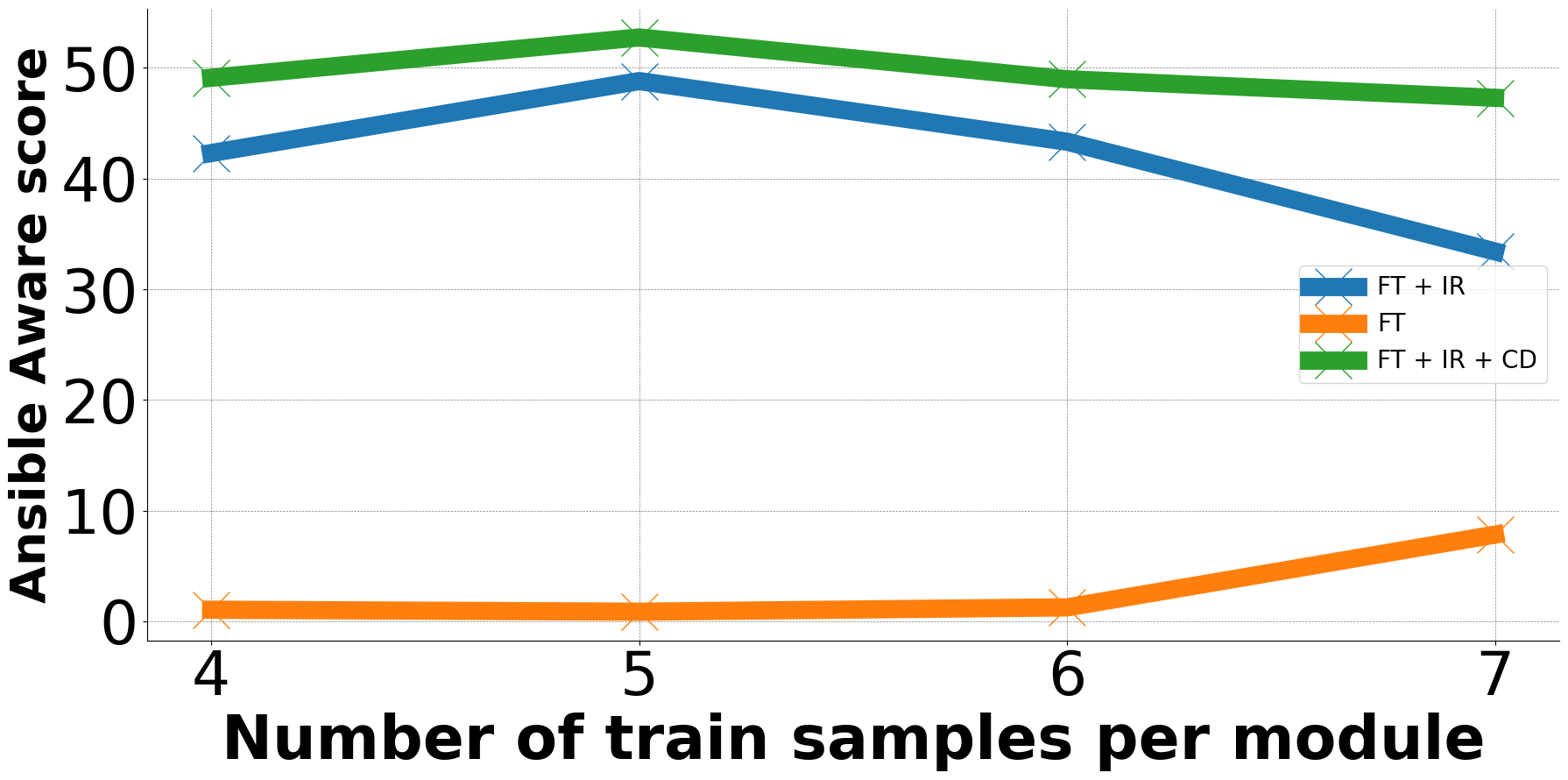} &   \includegraphics[width=50mm]{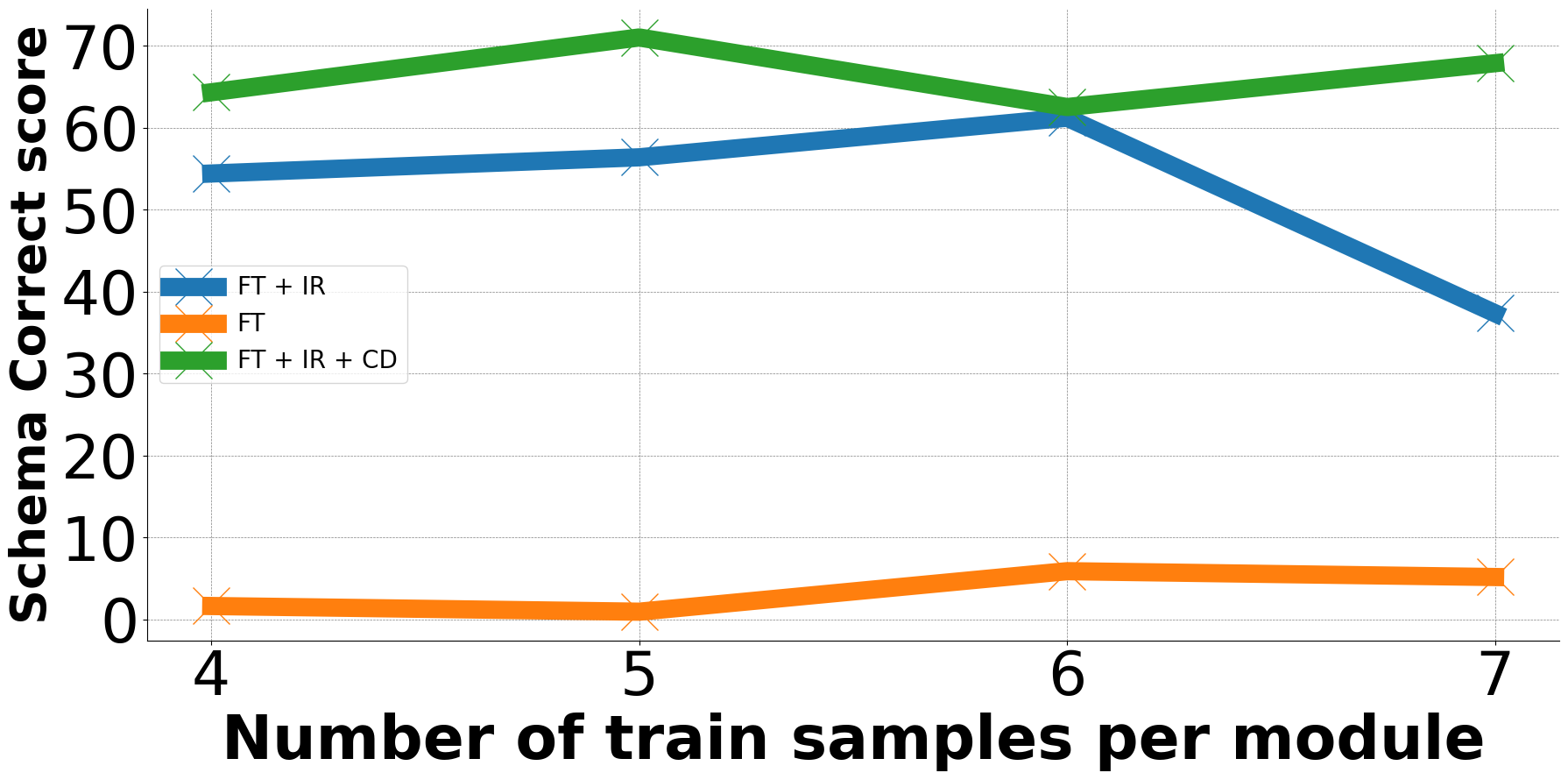} & \includegraphics[width=50mm]{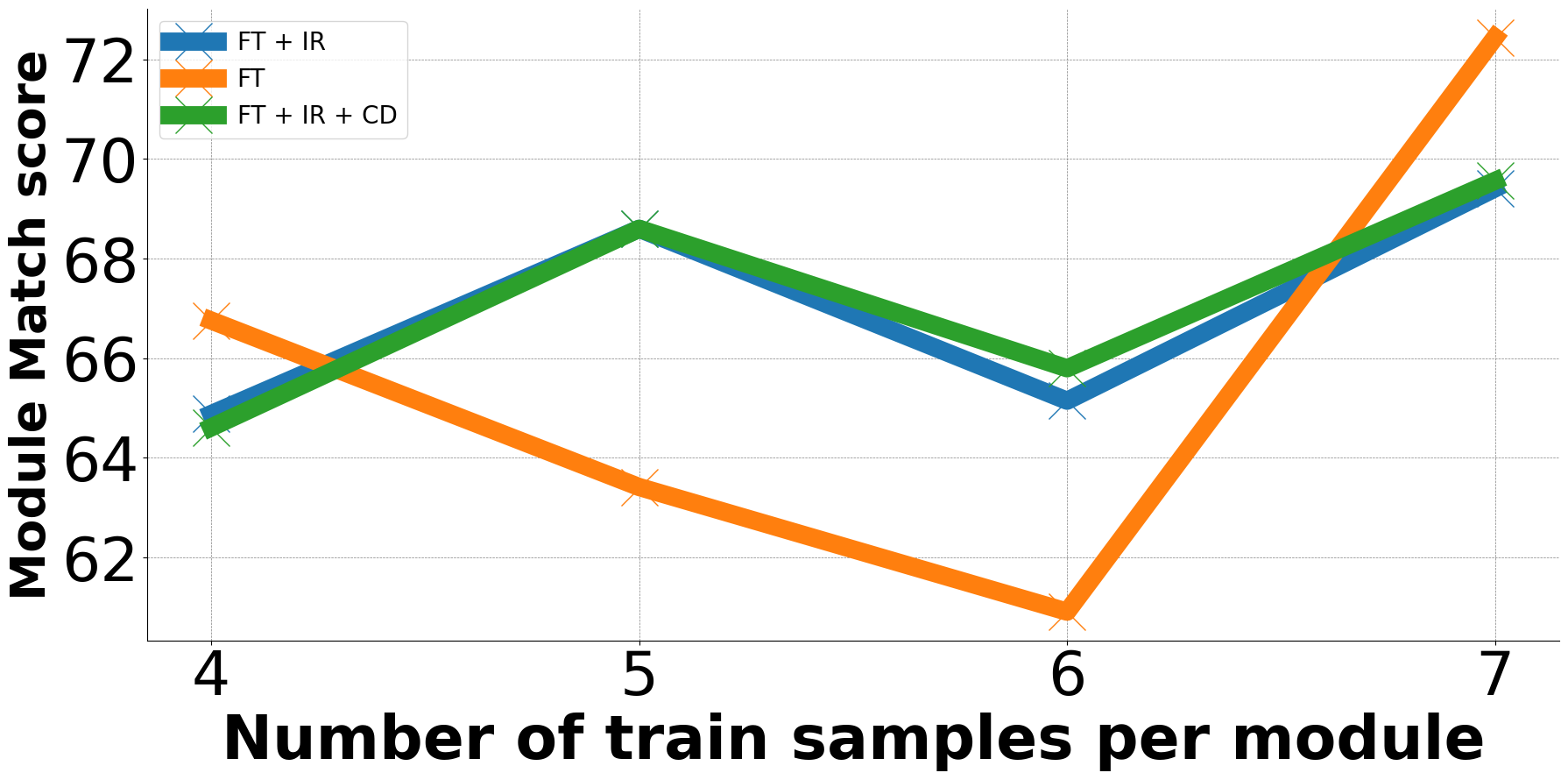}  \\
(d) & (e) & (f) \\
  \includegraphics[width=50mm]{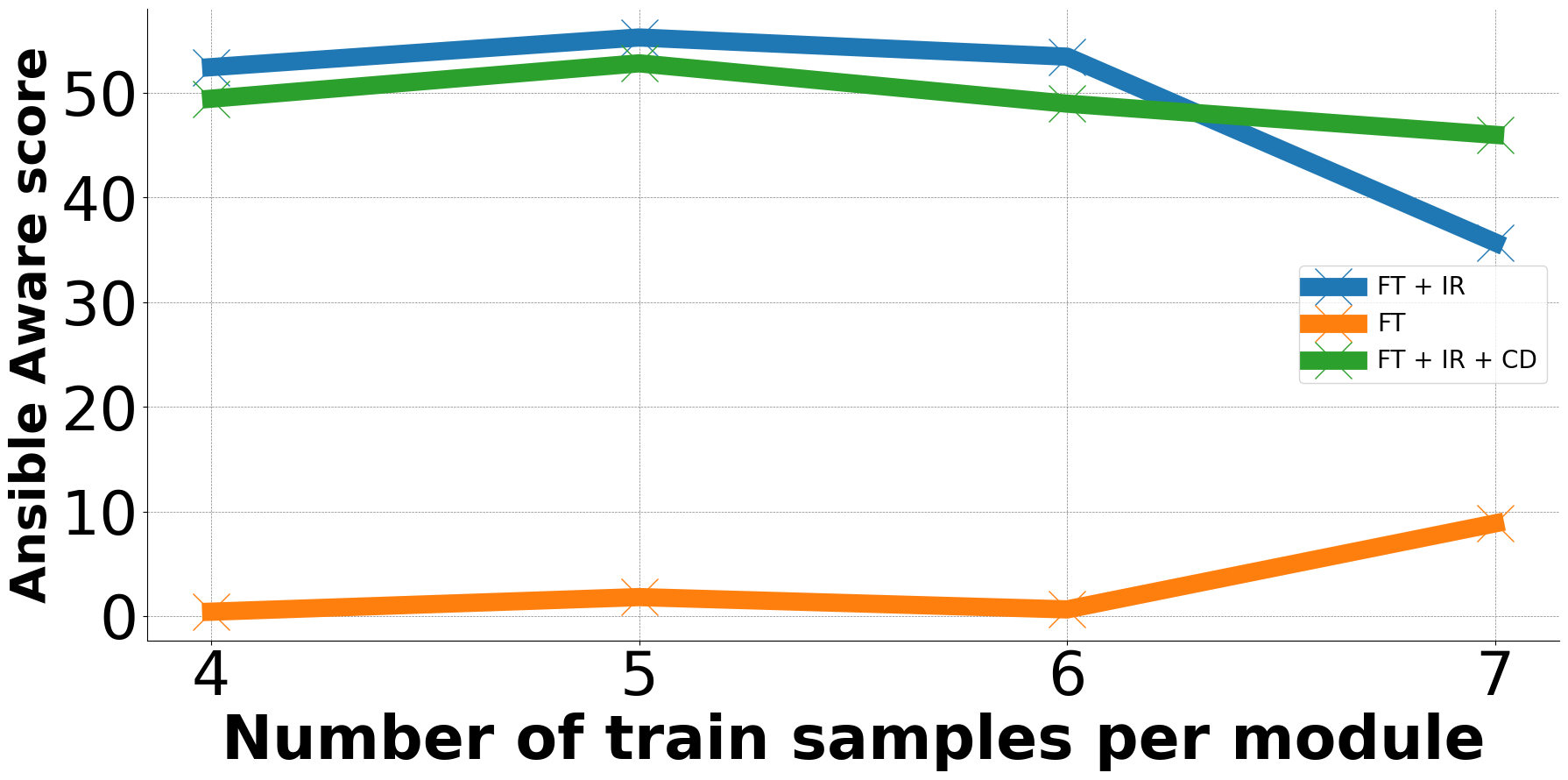} &   \includegraphics[width=50mm]{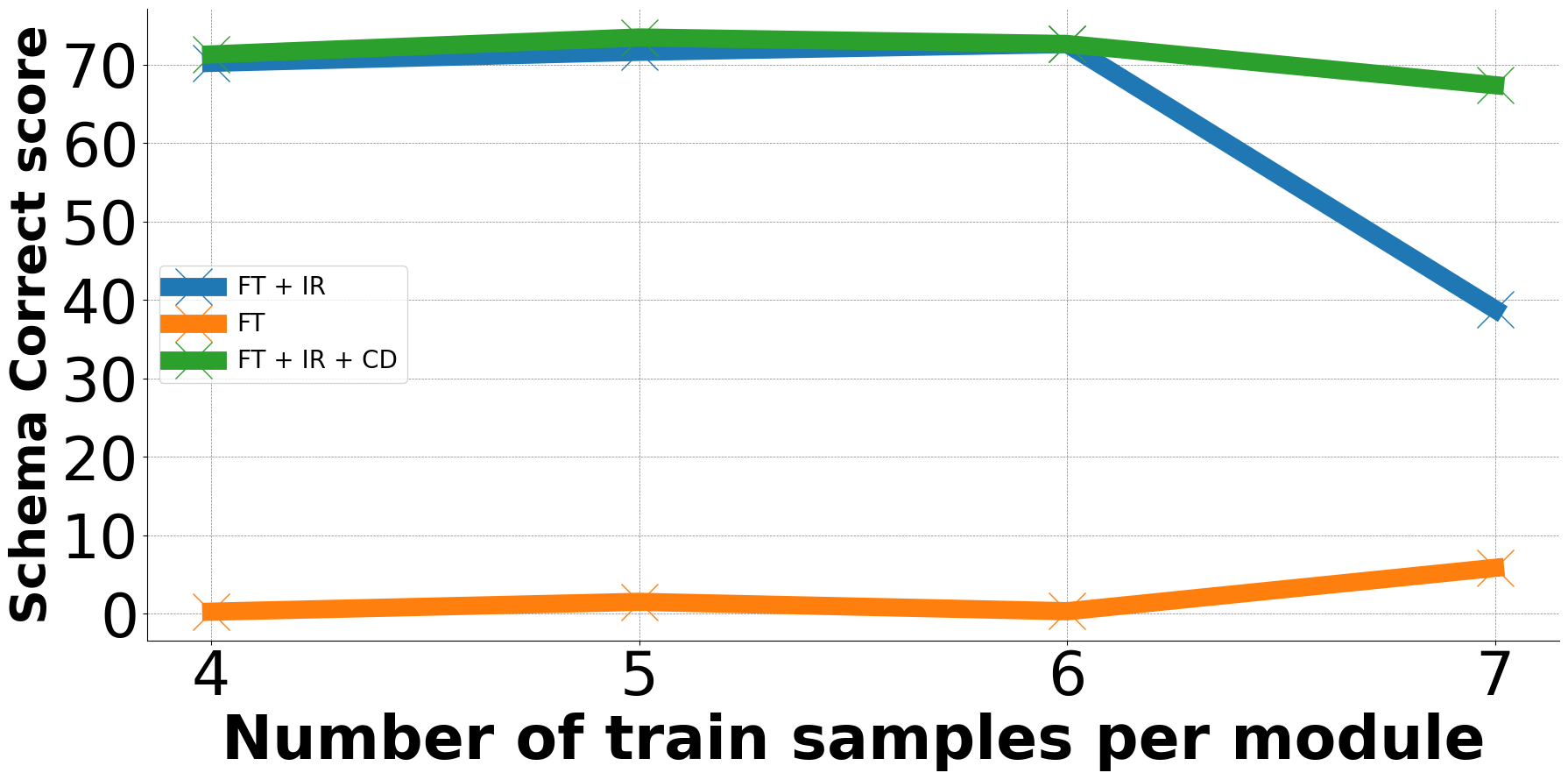} & \includegraphics[width=50mm]{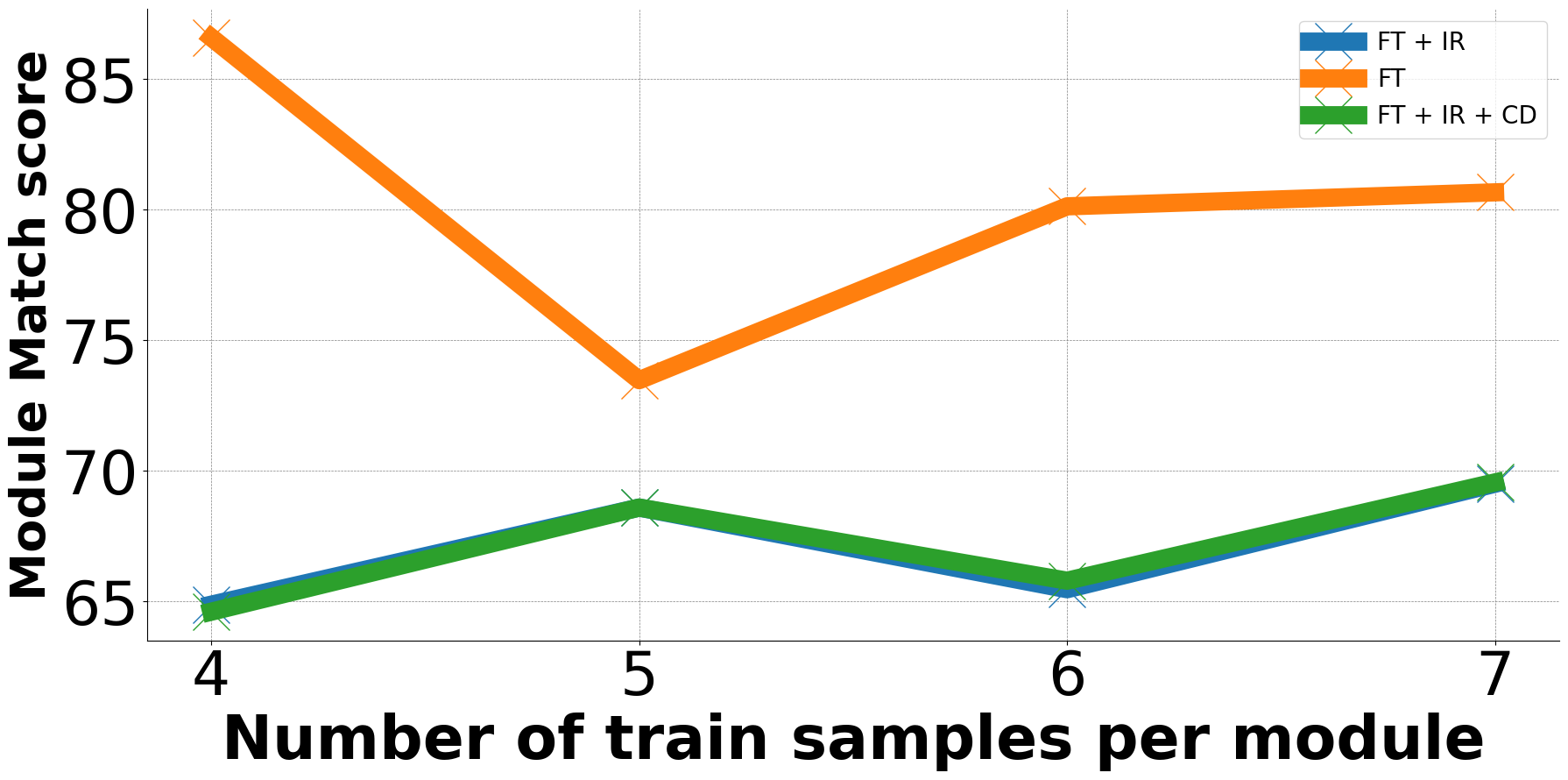}  \\
(g) & (h) & (i) \\
\end{tabular}
\caption{Demonstration of the performance of (a) (b) (c) GPT Neo 1.3B, (d) (e) (f) StarCoder2 3B, and (g) (h) (i) StarCoder2 7B in different configurations for \nlyaml{} task over varying number of train samples per module for in domain setting. We omit CodeLlama 34B as it is evaluated in few-shot setting.}
\label{plot:lowresource-other}
\end{figure*}

\begin{table*}[ht]
\centering
\begin{tabular}{ccllccclcllcll}
\toprule
\multirow{2}{*}{\small \textbf{Model}} &
  \multicolumn{5}{c}{\small \textbf{Bash}} &
  \multicolumn{8}{c}{\small \textbf{Ansible YAML}} \\ \cline{2-14} 
 &
  \multicolumn{3}{c}{\textbf{\begin{tabular}[c]{@{}c@{}}\small Exact \\ \small Match (\%)\end{tabular}}} &
  \textbf{\begin{tabular}[c]{@{}c@{}}\small CMD \\ \small Acc (\%)\end{tabular}} &
  \textbf{\small Token F1} &
  \multicolumn{2}{c}{\textbf{\begin{tabular}[c]{@{}c@{}}\small Module \\ \small Acc (\%)\end{tabular}}} &
  \multicolumn{3}{c}{\textbf{\begin{tabular}[c]{@{}c@{}}\small Schema \\ \small Correct\end{tabular}}} &
  \multicolumn{3}{c}{\textbf{\begin{tabular}[c]{@{}c@{}}\small Ansible \\ \small Aware\end{tabular}}} \\ \hline
\small Codellama 34B (3 shot) &
  \multicolumn{3}{c}{\small 13.2} &
  \small 32.4 &
  \small 21.8 &
  \multicolumn{2}{c}{\small 12.35} &
  \multicolumn{3}{c}{\small 20.33} &
  \multicolumn{3}{c}{\small 3.54} \\
\small + IR&
  \multicolumn{3}{c}{\small 16.71} &
  \small \textbf{38.32} &
  \small 26.49 &
  \multicolumn{2}{c}{\textbf{\small 36.38}} &
  \multicolumn{3}{c}{\small 13.18} &
  \multicolumn{3}{c}{\small 7.39} \\
\small + IR + CD &
  \multicolumn{3}{c}{\small \textbf{19.63}} &
  \small \small \textbf{38.32} &
  \small \small \textbf{29.71} &
  \multicolumn{2}{c}{\small \textbf{36.38}} &
  \multicolumn{3}{c}{\small \textbf{65.72}} &
  \multicolumn{3}{c}{\small \textbf{15.77}} \\ \hline
\small StarCoder2 15B (3 shot) &
  \multicolumn{3}{c}{\small 11.78} &
  \small 30.71 &
  \small 19.63 &
  \multicolumn{2}{c}{\small 11.06} &
  \multicolumn{3}{c}{\small 4.32} &
  \multicolumn{3}{c}{\small 0.53} \\
\small + IR &
  \multicolumn{3}{c}{\small 15.62} &
  \small \textbf{38.32} &
  \small 24.71 &
  \multicolumn{2}{c}{\small 36.38} &
  \multicolumn{3}{c}{\small 12.05} &
  \multicolumn{3}{c}{\small 3.40} \\
\small + IR + CD &
  \multicolumn{3}{c}{\small \textbf{18.19}} &
  \small \textbf{38.32} &
  \small \textbf{31.83} &
  \multicolumn{2}{c}{\small \textbf{36.38}} &
  \multicolumn{3}{c}{\small \textbf{66.04}} &
  \multicolumn{3}{c}{\small \textbf{20.78}} \\ \bottomrule
\end{tabular}
\caption{Results for in-context learning for out-of-domain setting with and without IR and constrained decoding. Here, the model is constrained to follow the Top-1 retrieved library template only. Hence, Command Acc and Module Acc, which detect the exact match of the library in generated code, depend only on IR and give the same scores for IR and IR+CD models.}
\label{OOD in context}
\end{table*}

\begin{table*}[ht]
\centering
\begin{tabular}{ccllccclcllcll}
\toprule
\multirow{2}{*}{\small \textbf{Model}} &
  \multicolumn{5}{c}{\small \textbf{Bash}} &
  \multicolumn{8}{c}{\small \textbf{Ansible YAML}} \\ \cline{2-14} 
 &
  \multicolumn{3}{c}{\textbf{\begin{tabular}[c]{@{}c@{}}\small Exact \\ \small Match (\%)\end{tabular}}} &
  \textbf{\begin{tabular}[c]{@{}c@{}}\small CMD \\ \small Acc (\%)\end{tabular}} &
  \textbf{\small Token F1} &
  \multicolumn{2}{c}{\textbf{\begin{tabular}[c]{@{}c@{}}\small Module \\ \small Acc (\%)\end{tabular}}} &
  \multicolumn{3}{c}{\textbf{\begin{tabular}[c]{@{}c@{}}\small Schema \\ \small Correct\end{tabular}}} &
  \multicolumn{3}{c}{\textbf{\begin{tabular}[c]{@{}c@{}}\small Ansible \\ \small Aware\end{tabular}}} \\ \hline
\small StarCoder2 3B &
  \multicolumn{3}{c}{\small 4.09} &
  \small 17.88 &
  \small 34.22 &
  \multicolumn{2}{c}{\small 25.12} &
  \multicolumn{3}{c}{\small 4.65} &
  \multicolumn{3}{c}{\small 5.35} \\
\small + IR (Top 3) + CD &
  \multicolumn{3}{c}{\small \textbf{5.24}} &
  \small \textbf{27.33} &
  \small \textbf{36.50} &
  \multicolumn{2}{c}{\textbf{\small 27.29}} &
  \multicolumn{3}{c}{\textbf{\small 49.45}} &
  \multicolumn{3}{c}{\textbf{\small 17.66}} \\
\small + IR (Top 10) + CD &
  \multicolumn{3}{c}{\small 4.88} &
  \small \small 25.31 &
  \small \small 34.91 &
  \multicolumn{2}{c}{\small 24.52} &
  \multicolumn{3}{c}{\small 47.8} &
  \multicolumn{3}{c}{\small 15.25} \\ \hline
\small StarCoder2 7B &
  \multicolumn{3}{c}{\small 4.12} &
  \small 16.16 &
  \small 34.45 &
  \multicolumn{2}{c}{\small 22.13} &
  \multicolumn{3}{c}{\small 5.16} &
  \multicolumn{3}{c}{\small 5.61} \\
\small + IR (Top 3) + CD &
  \multicolumn{3}{c}{\small \textbf{5.61}} &
  \small \textbf{26.41} &
  \small \textbf{37.71} &
  \multicolumn{2}{c}{\textbf{\small 25.41}} &
  \multicolumn{3}{c}{\textbf{\small 47.81}} &
  \multicolumn{3}{c}{\textbf{\small 19.32}} \\
\small + IR (Top 10)+ CD &
  \multicolumn{3}{c}{\small 4.31} &
  \small 24.14 &
  \small 33.73 &
  \multicolumn{2}{c}{\small 23.82} &
  \multicolumn{3}{c}{\small 45.62} &
  \multicolumn{3}{c}{\small 17.14} \\ \bottomrule
\end{tabular}
\caption{Results for each base fine-tuned language model for out-of-domain setting with and without IR (top 3 and 10 retrievals) and constrained decoding.}
\label{OOD top 3}
\end{table*}

\begin{table*}[ht]
\centering
\begin{tabular}{ccllccclcllcll}
\hline
\multirow{2}{*}{\small \textbf{Model}} &
  \multicolumn{5}{c}{\small \textbf{\small Bash}} &
  \multicolumn{8}{c}{\small \textbf{Ansible YAML}} \\ \cline{2-14} 
 &
  \multicolumn{3}{c}{\textbf{\begin{tabular}[c]{@{}c@{}}\small Exact \\ \small Match (\%)\end{tabular}}} &
  \textbf{\begin{tabular}[c]{@{}c@{}}\small CMD \\ \small Acc (\%)\end{tabular}} &
  \textbf{\small \small Token F1} &
  \multicolumn{2}{c}{\textbf{\begin{tabular}[c]{@{}c@{}}\small Module \\ Acc (\%)\end{tabular}}} &
  \multicolumn{3}{c}{\textbf{\begin{tabular}[c]{@{}c@{}}\small Schema \\ \small Correct\end{tabular}}} &
  \multicolumn{3}{c}{\textbf{\begin{tabular}[c]{@{}c@{}}\small Ansible \\ \small Aware\end{tabular}}} \\ \hline
\small StarCoder2 3B &
  \multicolumn{3}{c}{\small 15.26} &
  \small 47.91 &
  \small 50.38 &
  \multicolumn{2}{c}{\small 52.79} &
  \multicolumn{3}{c}{\small 4.65} &
  \multicolumn{3}{c}{\small 5.25} \\
\small + IR (Top 3) + CD &
  \multicolumn{3}{c}{\small \textbf{16.71}} &
  \small \textbf{54.55} &
  \small \textbf{54.31} &
  \multicolumn{2}{c}{\small \textbf{56.21}} &
  \multicolumn{3}{c}{\small \textbf{49.37}} &
  \multicolumn{3}{c}{\textbf{\small 36.21}} \\
\small + IR (Top 10) + CD &
  \multicolumn{3}{c}{\small 15.51} &
  \small 53.22 &
  \small 52.89 &
  \multicolumn{2}{c}{\small 46.62} &
  \multicolumn{3}{c}{\small 47.56} &
  \multicolumn{3}{c}{\small 34.24} \\ \hline
\small StarCoder2 7B &
  \multicolumn{3}{c}{\small 14.91} &
  \small 46.99 &
  \small 50.82 &
  \multicolumn{2}{c}{\textbf{\small 77.95}} &
  \multicolumn{3}{c}{\small 4.38} &
  \multicolumn{3}{c}{\small 6.49} \\
\small + IR (Top 3) + CD &
  \multicolumn{3}{c}{\small \textbf{16.27}} &
  \small \textbf{53.44} &
  \small \textbf{54.07} &
  \multicolumn{2}{c}{\small 58.56} &
  \multicolumn{3}{c}{\small \textbf{47.13}} &
  \multicolumn{3}{c}{\textbf{\small 33.51}} \\
\small + IR (Top 10)+ CD &
  \multicolumn{3}{c}{\small 15.22} &
  \small 51.15 &
  \small 52.49 &
  \multicolumn{2}{c}{\small 50.15} &
  \multicolumn{3}{c}{\small 45.38} &
  \multicolumn{3}{c}{\small 30.76} \\ \hline
\end{tabular}
\caption{Results for each base fine-tuned language model for in-domain setting with and without IR (top 3 and 10 retrievals) and constrained decoding.}
\label{Id top 3}
\end{table*}

\begin{table*}[ht]
\centering
\begin{tabular}{ccllccclcllcll}
\hline
\multirow{2}{*}{\small \textbf{Model}} &
  \multicolumn{5}{c}{\small \textbf{\small Bash}} &
  \multicolumn{8}{c}{\small \textbf{Ansible YAML}} \\ \cline{2-14} 
 &
  \multicolumn{3}{c}{\textbf{\begin{tabular}[c]{@{}c@{}}\small Exact \\ \small Match (\%)\end{tabular}}} &
  \textbf{\begin{tabular}[c]{@{}c@{}}\small CMD \\ \small Acc (\%)\end{tabular}} &
  \textbf{\small \small Token F1} &
  \multicolumn{2}{c}{\textbf{\begin{tabular}[c]{@{}c@{}}\small Module \\ Acc (\%)\end{tabular}}} &
  \multicolumn{3}{c}{\textbf{\begin{tabular}[c]{@{}c@{}}\small Schema \\ \small Correct\end{tabular}}} &
  \multicolumn{3}{c}{\textbf{\begin{tabular}[c]{@{}c@{}}\small Ansible \\ \small Aware\end{tabular}}} \\ \hline
\small StarCoder2 3B &
  \multicolumn{3}{c}{\small 4.18} &
  \small 17.13 &
  \small 32.78 &
  \multicolumn{2}{c}{\small 26.16} &
  \multicolumn{3}{c}{\small 4.96} &
  \multicolumn{3}{c}{\small 5.90} \\
\small + IR (Top 1) &
  \multicolumn{3}{c}{\small 5.12} &
  \small \textbf{38.32} &
  \small 39.81 &
  \multicolumn{2}{c}{\small 36.38} &
  \multicolumn{3}{c}{\small 22.47} &
  \multicolumn{3}{c}{\small 11.12} \\
\small + IR + CD &
  \multicolumn{3}{c}{\small \textbf{6.24}} &
  \small \textbf{38.32} &
  \small \textbf{41.73} &
  \multicolumn{2}{c}{\small \textbf{36.38}} &
  \multicolumn{3}{c}{\small \textbf{31.21}} &
  \multicolumn{3}{c}{\small \textbf{16.26}} \\ \hline
\small StarCoder2 7B &
  \multicolumn{3}{c}{\small 5.49} &
  \small 17.88 &
  \small 35.72 &
  \multicolumn{2}{c}{\small 21.98} &
  \multicolumn{3}{c}{\small 5.11} &
  \multicolumn{3}{c}{\small 5.63} \\
\small + IR (Top 1) &
  \multicolumn{3}{c}{\small \textbf{6.23}} &
  \small \textbf{38.32} &
  \small \textbf{40.71} &
  \multicolumn{2}{c}{\small \textbf{36.38}} &
  \multicolumn{3}{c}{\small 3.93} &
  \multicolumn{3}{c}{\small 3.23} \\
\small + IR + CD &
  \multicolumn{3}{c}{\small 7.81} &
  \small \textbf{38.32} &
  \small \textbf{42.31} &
  \multicolumn{2}{c}{\small \textbf{36.38}} &
  \multicolumn{3}{c}{\small \textbf{43.43}} &
  \multicolumn{3}{c}{\small \textbf{16.38}} \\ \hline
\end{tabular}
\caption{Results for each pre-trained and further fine-tuned language model for OOD setting with and without IR (top 1) and constrained decoding.}
\label{Pre-train OOD top 1}
\end{table*}

\begin{table*}[ht]
\centering
\begin{tabular}{ccllccclcllcll}
\hline
\multirow{2}{*}{\small \textbf{Model}} &
  \multicolumn{5}{c}{\small \textbf{\small Bash}} &
  \multicolumn{8}{c}{\small \textbf{Ansible YAML}} \\ \cline{2-14} 
 &
  \multicolumn{3}{c}{\textbf{\begin{tabular}[c]{@{}c@{}}\small Exact \\ \small Match (\%)\end{tabular}}} &
  \textbf{\begin{tabular}[c]{@{}c@{}}\small CMD \\ \small Acc (\%)\end{tabular}} &
  \textbf{\small \small Token F1} &
  \multicolumn{2}{c}{\textbf{\begin{tabular}[c]{@{}c@{}}\small Module \\ Acc (\%)\end{tabular}}} &
  \multicolumn{3}{c}{\textbf{\begin{tabular}[c]{@{}c@{}}\small Schema \\ \small Correct\end{tabular}}} &
  \multicolumn{3}{c}{\textbf{\begin{tabular}[c]{@{}c@{}}\small Ansible \\ \small Aware\end{tabular}}} \\ \hline
\small StarCoder2 3B &
  \multicolumn{3}{c}{\small 15.26} &
  \small 48.38 &
  \small 51.74 &
  \multicolumn{2}{c}{\small 53.90} &
  \multicolumn{3}{c}{\small 4.71} &
  \multicolumn{3}{c}{\small 6.20} \\
\small + IR (Top 1) &
  \multicolumn{3}{c}{\small 16.71} &
  \small \textbf{60.12} &
  \small 54.61 &
  \multicolumn{2}{c}{\small 68.45} &
  \multicolumn{3}{c}{\small 39.11} &
  \multicolumn{3}{c}{\small 35.41} \\
\small + IR + CD &
  \multicolumn{3}{c}{\small \textbf{17.81}} &
  \small \textbf{60.12} &
  \small \textbf{56.73} &
  \multicolumn{2}{c}{\small \textbf{68.45}} &
  \multicolumn{3}{c}{\small \textbf{48.41}} &
  \multicolumn{3}{c}{\small \textbf{38.98}} \\ \hline
\small StarCoder2 7B &
  \multicolumn{3}{c}{\small 15.63} &
  \small 48.38 &
  \small 52.73 &
  \multicolumn{2}{c}{\textbf{\small 77.81}} &
  \multicolumn{3}{c}{\small 4.1} &
  \multicolumn{3}{c}{\small 6.39} \\
\small + IR (Top 1) &
  \multicolumn{3}{c}{\small \textbf{16.21}} &
  \small \textbf{60.12} &
  \small \textbf{54.77} &
  \multicolumn{2}{c}{\small 68.45} &
  \multicolumn{3}{c}{\small 45.60} &
  \multicolumn{3}{c}{\small 40.61} \\
\small + IR + CD &
  \multicolumn{3}{c}{\small 15.22} &
  \small 60.12 &
  \small 52.49 &
  \multicolumn{2}{c}{\small 68.45} &
  \multicolumn{3}{c}{\small \textbf{52.09}} &
  \multicolumn{3}{c}{\small \textbf{42.66}} \\ \hline
\end{tabular}
\caption{Results for each pre-trained and further fine-tuned language model for in-domain setting with and without IR (top 1) and constrained decoding.}
\label{Pre-train Id top 1}
\end{table*}

\subsection{NL to Bash}\label{nl2bash}
\label{subsec:appendix-nl2bash}
This section describes specifics of techniques used for NL to Bash task.

\subsubsection{Module Description and Constraints}

The \tldr{} dataset is not equipped with fine-grained information such as module description and constraints. The dataset has a total of $1503$ bash utilities. 

\paragraph{Module Descriptions:} 
Document for every bash utility consists of utility descriptions and NL to Bash examples from corresponding bash utility. Details for both components are given below.

\noindent\textbf{Utility Description: }
We scrape the descriptions of each bash utility from \emph{DESCRIPTION} section of Linux man-pages\footnote{\url{https://manned.org/pkg/ubuntu-mantic}}. Empirically, we observe that the bash utility descriptions are redundant after the first 60 tokens. Therefore, we select the first 60 tokens from the descriptions. However, if the description is shorter than $30$ words, we use full documentation as the description. 

\noindent\textbf{Examples:}
For both ID and OOD settings, we augment descriptions of utilities from the train set with two to three NL to bash example pairs. These pairs are randomly sampled from the training corpus itself. For example, if the bash utility \emph{tar} is in the train set, its document is augmented with NL to bash pairs from the train set having utility as \emph{tar}. This ensures that none of the examples from the test set are present in the document. Since utilities in the OOD split test set are disjointed from the train set, documents for the utilities in the OOD split test set consist of only utility descriptions.


\begin{lstlisting}[language=bash, caption=Example templates for bash command curated using synopsis section in linux man page. Here fields within {[]} denotes optional fields and {[a|b|c]} denotes that one of the strings among from a{,} b or c has to be generated, label={lst:bashsyntax}]
cp [OPTION] {{SOURCE}} {{DIRECTORY}}

needrestart [-{{v|q}} | -n | -c <cfg> | -r <mode> | -f <fe> | -u <ui> | -{{b|p}} | -kl]

git rename-tag {{old-tag-name}} {{new-tag-name}}

lzop [ command ] [ options ] [ filename ...  ]

meson setup [ options ] [ build directory ] [ source directory ]

gh <command> <subcommand> [flags]
\end{lstlisting}

\paragraph{Structured schema: }\label{schema}We augment \tldr{} dataset with schema information for every bash utility. We crawl the Linux man pages of bash modules and collect the initial template $T$ of the bash command for each library from \emph{usage} or \emph{SYNOPSIS} section. Further, we collect the list of valid options and sub-commands for each bash utility. Schema information also includes inter-field dependency information, like a list of valid flags and arguments for every subcommand. For example, for the Linux command \emph{cp}, some of the valid options are \emph{-a, --archive}, \emph{-f, --force}, and \emph{-i, --interactive} are scraped from linux man page. 

\paragraph{Templates: }Along with options, we also scrape the syntax of bash modules mentioned under \emph{usage} section. In \emph{SYNOPSIS} section, it is standard practice that text enclosed within [] is optional, and the presence and position of that field in the command are not fixed. Text enclosed within <> must be produced at the position in the template. For the optional fields, we use language-specific trigger signals $G$. Examples of bash command templates are given in listing \ref{lst:bashsyntax}.

\paragraph{Trigger signals: }\label{Bash triggers}Trigger signals used for bash are as follows. If the model generates the token " --," we constrain the model from generating the string from valid doublehand flags. Similar constraints are used for shorthand flags " -". Other trigger signals include the generation of a pipe operator ("|"). In the bash command, the pipe operator forwards the output of one process to another as input. For example, bash command \emph{nl -s prefix file.txt | cut -c7-} consists of two bash utilities \emph{nl} and \emph{cut} separated by "|". Generation of token "|" denotes the start of a new process with a new bash utility. Hence, while decoding, if the model generates an operator-like token (“|”), then we constrain the model to freshly follow one of the k templates from the start using the library selection algorithm again \ref{library selection}. This trigger signal allows us to generate the bash command with multiple utilities or processes.

\paragraph{Enforced schema rules: }We ensure that all the required fields (flags and subcommands) are generated according to their position specified in the template. Further, it is also ensured that all the generated flags and subcommands adhere to the library schema. For the templates that specify the compulsory arguments, we treat those arguments as \emph{static} part of the template and include it in the final output. For example, as given in the template of bash utility \emph{cp}, \emph{source} and \emph{directory} are the compulsory arguments and hence directly included in the output command.

\subsection{Pre-training data}\label{pretrain-bash}
We append the Linux man-pages for $1.5k$ bash utilities in a single file which is used for pre-training\footnote{\url{https://manned.org/}}. For every man page, we remove all newline characters and replace double newline characters with a single newline. This keeps the definition of each flag and field separate from each other and results in better performance. The final pre-training data consists of $10.3$ million tokens.

\subsection{Hyperparameter details}
We use NVIDIA A100 $80$ GB GPUs to perform inference and training for all the experiments. We use the standard HuggingFace transformers \cite{wolf-etal-2020-transformers} with accelerate to load, train, and perform inference for all the models. For constrained decoding we use HuggingFace logits processor\footnote{\url{https://huggingface.co/docs/transformers/en/internal/generation_utils\#transformers.LogitsProcessor}}.

\subsubsection{Ansible YAML}
All fine-tuned models are fully parameter-tuned to the task. For fine-tuning, we used Adam optimizer with batch size two for all the models and context length of $2048$. We also use the linear learning scheduler and a learning rate of $4e-5$. At inference, we experimented with both greedy search and beam search-based decoding techniques for baselines, and we observed beam search with $5$ number of beams performed the best. Training is done for two epochs. All the models are used in bf16 precision.

We use the bert-based-uncased model as base and fine-tune the standard ColBERTv2 pre-trained model\footnote{\url{https://github.com/stanford-futuredata/ColBERT}} on \nlyaml{} task. The document corpus size is $2922$ documents. We run the fine-tuning task for $5000$ max number of steps. We use $8$ negatives for every query while preparing the triplets. The train-test splits for fine-tuning follow the numbers from language model fine-tuning (Table \ref{table:data-stats}).

\subsubsection{Bash command}
All the training details for bash command generation are the same as those for ansible YAML, except that we use a batch size of $4$ with gradient accumulation steps of $4$ during fine-tuning. The maximum sequence length for the bash command is $512$. All the models are used here in fp32 precision.

Similar to \nlyaml{} task, we use the pre-trained ColBERTv2 for fine-tuning the task data. The document corpus size is $1503$ documents. Similar to \nlyaml{} task, we run for a max of $5000$ number of steps. We use $8$ negatives for every query while preparing the triplets.

\subsubsection{Pre-training}
For pre-train the language models on the next word prediction task using library documentation for $3$ epochs. For pre-training we use a cosine scheduler with a learning rate of $5e-05$. We experiment with both linear and cosine schedulers and use cosine scheduler checkpoints for further fine-tuning due to the best results. We pre-train with a batch size of 4, gradient accumulation steps of 8, and bf16 precision. Due to scarce data, we use warmup steps of 100 for bash and 150 for ansible pre-training. We use the block size of 1024 for pre-training.

\label{subsec:appendix-modeltraininf}

\subsection{Analysis}

\paragraph{Promising low data resource performance:} First, DocCGen outperforms all the baselines in the OOD setting (Table \ref{table:OOD}) and performs competitively across overall degrees of low-resource data (Figure \ref{plot:lowresource}) in ID setting.  Second, the performance of fine-tuned StarCoder2 3B in generating good YAML code following the ansible module improves gradually for \ansbileaware{} and \schemacorrect{} metrics with an increase in training samples. However, extrapolating this growth to meet DocCGen's performance might require a large number of training samples per module. Third, DocCGen outperforms baselines in most of the lower orders of training sample count for \modulematch{} metric. This behavior is consistent across all models. (Figure \ref{plot:lowresource-other})

\begin{table*}[ht]
\centering
\begin{tabular}{cccc}
\toprule
\multirow{2}{*}{\textbf{Model}} & \multicolumn{3}{c}{\textbf{Bash}} \\ \cline{2-4} 
 & \textbf{\begin{tabular}[c]{@{}c@{}}Template\\ Match (\%)\end{tabular}} & \textbf{\begin{tabular}[c]{@{}c@{}}Command \\ Acc (\%)\end{tabular}} & \textbf{Token F1} \\ \hline
StarCoder 1B                    & \small 14.32         & \small 57.34         & \small 58.42         \\
+ IR + CD                       & \small \textbf{18.92}         & \small \textbf{73.24}         & \small \textbf{66.47}         \\ \hline
StarCoder 3B                    & \small 16.34         & \small 61.34         & \small 62.34         \\
+ IR + CD                       & \small \textbf{18.39}         & \small \textbf{73.87}         & \small \textbf{66.89}         \\ \bottomrule
\end{tabular}
\caption{Results for NL2bash dataset using Top-1 IR}
\label{NL2bash}
\end{table*}




\end{document}